\documentclass[12pt]{article}
\def\d{{\rm d}}
\usepackage{authblk}
\usepackage{blindtext}
\usepackage[utf8]{inputenc}
\usepackage{amsmath}
\usepackage{amsfonts}
\usepackage{amssymb}
\usepackage{graphicx}
\usepackage{caption}
\usepackage{subcaption}

\author[1,2]{R. Thiru Senthil\thanks{rtsenthil@imsc.res.in, https://orcid.org/0000-0002-7142-4527}}
\author[1,2]{D. Indumathi\thanks{indu@imsc.res.in, https://orcid.org/0000-0001-6685-4760}}
\author[2,3]{Prashant Shukla\thanks{prashant.shukla@gmail.com, https://orcid.org/0000-0001-8118-5331}}
\affil[1]{The Institute of Mathematical Sciences, Taramani, Chennai 600113, India}
\affil[2]{Homi Bhabha National Institute, Anushakti Nagar, Mumbai 400094, India}
\affil[3]{Nuclear Physics Division, Bhabha Atomic Research Centre, Mumbai 400085, India}
\date{\today}
\title{
{~ \hfill {\small \it IMSc/2022/01/01}} \\ [1.5cm]
A simulation study of tau neutrino events at the ICAL detector in INO}
\begin{document}
\maketitle

\paragraph{Abstract}: We present the first detailed simulation study
of tau neutrino-induced charged current (CC) events from atmospheric
neutrino interactions in the Iron Calorimeter (ICAL) detector at the
proposed India-based Neutrino Observatory (INO) laboratory. Since the
intrinsic atmospheric neutrino flux at few to 10s of GeV energy comprises
only electron and muon neutrinos (and anti-neutrinos) with negligible tau
neutrino component, any signature of atmospheric tau neutrinos is a signal
for neutrino oscillations. We study the tau leptons produced through
these CC interactions via their hadronic decay. These events appear as
an excess over the neutral current (NC) background where hadrons are the
only observable component. We find that the presence of tau neutrinos in
the atmospheric neutrino flux can be demonstrated to nearly $4\sigma$
confidence with 10 years data; in addition, these events are sensitive
to the neutrino oscillation parameters, $\sin^2\theta_{23}$ and $\vert
\Delta m_{31}^2 \vert$ (or $\vert \Delta m_{32}^2 \vert$), in the
2--3 sector. Finally, we show that combining these events with the
standard muon analysis which is the core goal of ICAL further improves
the precision with which these parameters, especially the octant of
$\theta_{23}$, can be measured.



\section{Introduction}
The proposed deep underground facility for carrying out research in basic
sciences, the India-based Neutrino Observatory (INO), will house the
magnetised Iron Calorimeter (ICAL) detector for carrying out atmospheric
neutrino experiments. The ICAL detector will consist of 51 ktons of
magnetised iron arranged in 151 layers, interspersed with Resistive Plate
Chambers (RPCs) as the active detector. The experiment aims to precisely
determine some of the important neutrino oscillation parameters using
atmospheric electron and muon neutrinos (and anti-neutrinos) as sources
\cite{ICAL:2015stm}.

The central goal of the proposed magnetised ICAL detector is the study
of the neutrino mass ordering/hierarchy through the separate detection of
muons and anti-muons (so-called standard muon events) produced in
charged-current (CC) interactions of atmospheric muon neutrinos and
anti-neutrinos, taking advantage of the magnetic field in ICAL. The
detector will be optimised to detect muons (and associated hadrons)
produced in the quasi-elastic, resonant, or deeply inelastic interactions
of neutrinos with few to 10s of GeV energy. Such a detector has been
shown, through detailed simulations, to be capable
of making precision measurements of the 2--3 oscillation parameters
$\sin^2\theta_{23}$ and $\Delta m^2_{32}$, while being {\em completely
insensitive} to the CP phase, $\delta_{CP}$ \cite{ICAL:2015stm}.
It is also sensitive, due to Earth matter effects and the ability to
identify the charge of muons, to the important open question of the
neutrino mass ordering. The detector also has (reduced) sensitivity
to these oscillation parameters through measurement of the electrons
(and positrons) produced in CC interactions of electron neutrinos with
the detector \cite{Chacko:2019wwm}.

In this paper, we explore the other conventional physics that can
be addressed by such a detector and study tau lepton production
through atmospheric neutrinos in particular. Studies of sensitivity to
electrons, as well as other exotic possibilities such as CPT/Lorentz
invariance violation, neutrino decay, etc., have been studied by
various members of the collaboration and the main results can be found in
Ref.~\cite{ICAL:2015stm}. Here we present the sensitivity to the presence
of tau neutrinos in the atmospheric neutrino flux that arise purely
from oscillations of $\nu_e$ and $\nu_\mu$ (and their anti-particles)
in the atmospheric neutrino fluxes.

The Super Kamiokande collaboration has analysed their data for signatures
of tau neutrinos and found that a no-tau hypothesis is rejected at the
$3.8\sigma$ level \cite{Super-Kamiokande:2012xtd}. They also determined
the overall normalisation of the tau neutrino flux. Recently, the
IceCube Collaboration has shown the presence of tau neutrinos arising from
neutrino oscillations in atmospheric neutrinos to $3\sigma$ confidence and
has also analysed the data to determine some of the neutrino oscillation
parameters \cite{IceCube:2019dqi}. In fact, it was pointed out in
Ref.~\cite{Denton:2021rsa} that the presence of tau neutrinos can be
established at IceCube without knowing the oscillation parameters, and
also without any assumption about the unitarity of the PMNS mixing
matrix. Tau neutrino signatures are also being explored
by the DUNE \cite{DeGouvea:2019kea} and KM3NeT/ORCA experiments
\cite{KM3NeT:2021ozk}. We show here that the presence of tau neutrinos
in the atmospheric neutrino flux can be established to nearly $4\sigma$
confidence with 10 years data at ICAL. Moreover, we find that the tau
sample (although contaminated with the neutral current (NC) sample) is
indeed sensitive to the 2--3 parameters $\sin^2\theta_{23}$ and $\Delta
m_{32}^2$; in addition, we show that a combined study of tau neutrino
events with the standard muon events will significantly improve the
precision with which these parameters can be measured at ICAL.

\section{Interaction of tau neutrinos with the detector}

Atmospheric neutrinos are produced from the primary and secondary
interactions of cosmic rays with Earth's atmosphere and comprise
electron and muon neutrinos and anti-neutrinos. The majority of these
interaction processes occur within a height of approximately 15 km
from the Earth's surface. Since tau neutrinos arise from the production
and decays of $D_S^\pm$ mesons in contrast to electron and muon neutrinos
which arise from the decays of the lighter $\pi^\pm$ pions, there is a
negligible intrinsic tau neutrino flux in the atmosphere at energies
less than 100 GeV. In fact, it has been shown that the intrinsic neutrino
fluxes produced in the Earth's atmosphere via $p\,p$ interactions are
present in the ratio $\nu_e:\nu_\mu:\nu_\tau::1:2:3\times 10^{-5}$
\cite{Athar:2005wg}. Since muon neutrinos
arise both from the decay of the pion component of cosmic rays as well
as from the subsequent decays of the muons produced in these decays,
the ratio of $\nu_e:\nu_\mu$ in atmospheric neutrinos is approximately
$1:2$ in the few GeV energy range. When these atmospheric neutrinos pass
through the Earth to reach the detector, neutrino flavour oscillations
occur, and tau neutrinos can arise via oscillations of both electron-
and muon-type neutrinos. Consequently, the atmospheric neutrino
species get re-distributed in the ratio $1:1:1$, provided the ratio
of the path length traversed to their energy, $L (\hbox{km})/E_\nu
(\hbox{GeV}) \gtrsim 330$ \cite{Pasquali:1998xf,Athar:2001jw,Lee:2004zm}.
For neutrinos in the 10s of GeV range, this holds for the neutrinos
entering the detector from below, or the so-called up-going neutrinos.
Hence the atmospheric neutrino fluxes contain a significant fraction
of $\nu_\tau$ in the upward direction (and practically none in the
downward direction).

In principle, these tau neutrinos can be detected via their charged
current (CC) interaction with the nucleons in the detector that produce
charged $\tau$ leptons and hadrons via
\begin{equation}
\nu_{\tau} \, N \rightarrow \tau^- \, X~; \hspace{1cm}
\overline{\nu}_{\tau} \, N \rightarrow \tau^+ \, X~, 
\end{equation}
where $X$ are hadrons (containing at least one nucleon to conserve
baryon number). These $\tau$ leptons will decay promptly, primarily into
hadrons. The branching fractions in the leptonic and hadronic channels are
\begin{eqnarray} \nonumber
B(\tau^- \rightarrow e^{-} \bar{\nu_e} \nu_{\tau}) & \sim & 17\%~, \\ \nonumber
B(\tau^- \rightarrow \mu^{-} \bar{\nu_{\mu}} \nu_{\tau}) & \sim & 17\% ~, \\
B(\tau^- \rightarrow \nu_{\tau} H') & \sim & 66\% ~,
\label{eq:decay}
\end{eqnarray}
with similar fractions for $\tau^+$ as well. It can be seen from the
last expression in Eq.~\ref{eq:decay} that the dominant branching in the
hadronic mode gives rise to additional hadrons, $H'$. The total energy in
hadrons in such charged current interactions of tau neutrinos therefore
arises from the contributions from $X$ and $H'$, that is, from both primary
production as well as subsequent decay. Hence it is expected that CC-tau
events will present as events with high hadronic energy and no observed
final state lepton. Due to the kinematics of the tau production and
decay process, these hadrons are also peaked in the direction of the
incident neutrino.

In addition to CC interactions, Neutral Current (NC) interactions from
all flavours of neutrinos also produce hadrons ($X'$) via
\begin{equation}
\nu_i \, N \rightarrow \nu_i \, X'~; ~~~i = e, \mu, \tau~.
\end{equation}
These cannot be distinguished from the CC-tau events, {\it i.e.}, hadrons
($X + H'$) due to CC interactions of $\nu_\tau$ and hadrons ($X'$) due to
NC interactions of all neutrino flavours, cannot be separated. Therefore,
the NC events act as an inseparable background to the CC-tau events. Hence
we will study the combined sample of all NC and CC-tau events in what
follows.

Note that the muon events arising from CC interactions of muon neutrinos
(and anti-neutrinos) in ICAL give rise to events with a characteristic
muon track associated with the hadron shower and hence can be
distinguished from the NC or CC tau events which are present only as hadronic
showers. Electron events from CC interactions of electron neutrinos
(and anti-neutrinos) also present as showers since the electromagnetic
shower of the electron and the hadronic shower from the associated
hadrons cannot be separated in ICAL. Such showers can be distinguished
from purely hadronic showers due to the larger number of hits per layer
\cite{Chacko:2019wwm}; hence in the present analysis we assume that the NC
and tau CC events can be (fully) separated from the CC electron and muon
events although not from each other.

Relevant neutrino energies in our study are $E_\nu > 3.5$ GeV (which is
the threshold for CC-tau production due to the large mass of the
$\tau$). The processes involved are thus predominantly in the deep
inelastic scattering (DIS) region. In this work, we have used the NUANCE
neutrino generator to generate events for CC-tau
and NC events in a simulated ICAL detector \cite{Casper:2002sd}. We
use a GEANT4-based simulation toolkit to study the response of
hadrons in ICAL with respect to energy and direction reconstruction
\cite{GEANT4:2002zbu}. Using this, we study the sensitivity of $\nu_\tau$
events (this henceforth refers to the combined CC-tau $+$ NC sample)
to the neutrino oscillation parameters. While these events are indeed
sensitive to these parameters, a vast improvement is found on combining
the tau sample with the standard CC muon sample. That will be discussed
in the following sections.

\section{Generation of CC-tau and NC events in ICAL}

We use the NUANCE neutrino generator to generate
the unoscillated events arising from both $\nu_e$ and $\nu_\mu$ fluxes
in the atmosphere. We then oscillate them in a 3-flavour model using the
Pontecorvo-Maki-Nakagawa-Sakata mixing matrix ($U$). Detector-dependent
resolutions and characteristics are then folded into the distributions.
The events are analysed for their sensitivity to the neutrino oscillation
parameters. We begin by decribing the generation of unoscillated events.

\subsection{Generation of unoscillated events using the NUANCE neutrino
generator}

The NUANCE neutrino generator was used to generate CC-tau events
with atmospheric muon and electron neutrinos \cite{Casper:2002sd}. The
HONDA3d fluxes were used to generate events for 1000 years of exposure at the
ICAL detector \cite{Honda:2006qj,Honda:2011nf,Honda:2019ymh}. The
distribution of tau events produced in the detector due to the interaction
of these tau neutrinos with the material (mostly iron) of the detector
is given by
\begin{equation}
\frac{\d^{2}N_\tau}{\d E_{\tau} \d\cos \theta_{\tau}} = T \times N_{D}
	\times \int \d E_\nu \d\cos\theta_\nu \d \phi_\nu \left(P_{\mu\tau} \frac{\d^{2} \Phi_{\mu}}
	{\d E_{\nu} \d \cos \theta_{\nu} \d\phi_\nu} +
	P_{e\tau} \frac{\d^{2} \Phi_{e}}
	{\d E_{\nu} \d \cos \theta_{\nu} \d\phi_\nu} \right)
\times \d^2 \sigma^{CC}_\tau~.
\label{eq:CCtau}
\end{equation}
A similar expression holds for tau anti-neutrinos as well. Here, $T$
is the exposure time in seconds, $N_{D}$ the total number of targets in the
detector, $\Phi_{e,\mu}$ the electron- and muon-type atmospheric neutrino
fluxes, $P_{i\tau}$ the relevant oscillation probabilities from flavours
$i \to \tau$, and $\sigma^{CC}_{\tau}$ the CC cross section for $\nu_\tau$
interactions. Note that the oscillation probabilities are independent of
the azimuthal angle and depend on $(E_\nu, \cos\theta_\nu)$ alone.
In addition, the differential cross section, given by
\begin{equation}
\d^2 \sigma^{CC}_\tau \equiv \frac{\d^2\sigma (E_\nu)} {\d E_\tau
\d\cos\theta_\tau}~,
\end{equation}
is used to obtain events in terms of the final lepton energy and angle.
The NUANCE generator further fragments the final baryon into hadrons
and lists the events in terms of the neutrino parameters $(E_\nu,
\cos\theta_\nu, \phi_\nu)$ as well as the individual parameters for each
particle in the final state including the lepton: $(E_f, \cos\theta_f,
\phi_f)$. When the final lepton is tau, it also completes the decay of the
tau in either the hadronic or semi-leptonic modes and lists these as
well. ``Unoscillated'' CC-tau events were generated assuming the $\nu_\mu$
and $\nu_e$ atmospheric fluxes to be $\nu_\tau$ fluxes and using the
CC-tau cross sections (for quasi-elastic, resonance and deep-inelastic
processes) coded into NUANCE, that is, the events corresponding
to the two terms in Eq.~\ref{eq:CCtau} were generated, excluding the
oscillation probability. These unoscillated events are weighted with
the appropriate oscillation probability during ``data'' generation
and analysis.

A similar procedure was used to generate the neutral current (NC) events
using NUANCE. Since NC events are insensitive to oscillations, the NC
events are simply generated from the $\nu_e$ and $\nu_\mu$ atmospheric
neutrino fluxes, using the NC cross sections instead of the CC cross
sections shown in Eq.~\ref{eq:CCtau} and without including any oscillation
probabilities. Both the CC-tau\footnote{In the entire analysis, we always
refer to the CC-tau events that decay hadronically. The 17\% each of CC tau
events that decay semi-leptonically producing muons or electrons will
add (insignificantly) to the CC muon or electron events arising from the
direct interaction of muon and electron neutrinos with the detector.}
and NC events give rise to events in the ICAL detector with hadron showers
alone and no charged lepton track. We now discuss the implementation of
neutrino oscillations.

\subsection{Neutrino oscillation parameters}

In the model with 3-flavour neutrino oscillation, the
Pontecorvo-Maki-Nakagawa-Sakata mixing matrix ($U$) is commonly
parametrised with three mixing angles $\theta_{12}, \theta_{13},
\theta_{23}$ and one charge parity violating phase ($\delta_{CP}$).
If the neutrinos are Majorana particles there exist two additional
Majorana CP phases; these are not visible in neutrino flavour
oscillations. The neutrino flavour oscillation probabilities depend in
general on the neutrino energy ($E_\nu$),
propagation distance between the source and the detector ($L$), parameters
of the oscillation matrix ($U$) and the mass squared differences
($\Delta m^{2}_{ij} = m^{2}_{i}-m^{2}_{j},~i\neq j$)
\cite{Wolfenstein:1977ue}. There are only
two independent mass square differences to be considered for the case
of three flavour neutrino oscillations.

The central values of the neutrino oscillation parameters and their
$3\sigma$ ranges used in this analysis are given in Table
\ref{tab:osc}. While the 1--2 parameters are kept fixed, the CP phase is
poorly known \cite{ParticleDataGroup:2020ssz}; moreover, the ICAL
experiment is not sensitive to this phase. Hence we fix $\delta_{CP}=0$
throughout this analysis. Since the octant of $\theta_{23}$ is still
unknown, its central value has been taken to be $\theta_{23}=45^\circ$
($\sin^2\theta_{23} = 0.5$) in this analysis, unless otherwise specified.

The as-yet unknown neutrino mass hierarchy depends on the neutrino mass
ordering:
\begin{eqnarray} \nonumber 
\Delta m_{31}^{2} & > & 0 \hbox{  for Normal Ordering}~, \\
\Delta m_{31}^{2} & < & 0 \hbox{  for Inverted Ordering}~.
\end{eqnarray}
Since $\Delta m_{21}^2 >0$  and $\vert \Delta m_{21}^2 \vert \ll \vert
\Delta m_{31}^2 \vert$, this means that $\Delta m_{32}^2$ and $\Delta
m_{31}^2$ have the same sign. For convenience, we define
\cite{Indumathi:2006gr}
\begin{equation}
\Delta m^{2} \equiv m_{3}^{2}-\left(\frac{m_{2}^{2}+m_{1}^{2}}{2}\right)~.
\label{eq:dm2}
\end{equation}
Note that $\Delta m^2$ flips sign {\em without changing its magnitude}
when the hierarchy/ordering changes and hence is a convenient parameter
compared to $\Delta m_{31}^2$ or $\Delta m_{32}^2$, which change in both
sign and magnitude depending on the mass ordering. We shall use $\Delta
m^2$ throughout in the analysis. Depending on the mass ordering, and using
the value of $\Delta m_{21}^2$ given in Table \ref{tab:osc}, the values
of $\Delta m_{31}^2$ and $\Delta m_{32}^2$ can be found from $\Delta
m^2$. We shall assume the normal ordering throughout this analysis,
unless otherwise specified. More details and the current values and ranges
of these parameters for both the normal and inverted mass ordering can
be found in Refs.~\cite{ParticleDataGroup:2020ssz,Esteban:2020cvm}.

\begin{table}[htp]
\centering
\vspace{0.1cm}
\begin{tabular}{| c | c | c |}
\hline
 Parameter & Central values & $3 \sigma $ Range \\ \hline
$\sin^2 \theta_{12}$ & $0.304$ & fixed \\
$\sin^2 \theta_{13}$ &$0.0222$ & $0.0203 \leftrightarrow 0.0241$\\
$\sin^2 \theta_{23}$ & $0.5$ & $0.381 \leftrightarrow 0.615$ \\
$\Delta m^2_{21}$ $(\times 10^{-5}$ eV$^2$) & $7.42$ & fixed \\
$\vert \Delta m^2 \vert$ $(\times 10^{-3}$ eV$^2$) & $2.47$ & $2.395 \leftrightarrow
2.564$  \\
$\delta_{CP} (^\circ)$ & $0.0$ & fixed \\ \hline 
\end{tabular}
\caption{The $3\sigma$ ranges of neutrino oscillation parameters --- mixing
angles and mass squared differences --- and central values used in the
present work \cite{ParticleDataGroup:2020ssz,Esteban:2020cvm}; $\Delta
m^2$ is defined in Eq.~\ref{eq:dm2}.}
\label{tab:osc}
\end{table}

\subsection{Neutrino oscillation probabilities}

\begin{figure}
     \centering
     \begin{subfigure}[t]{0.49\textwidth}
         \centering
         \includegraphics[width=\textwidth]{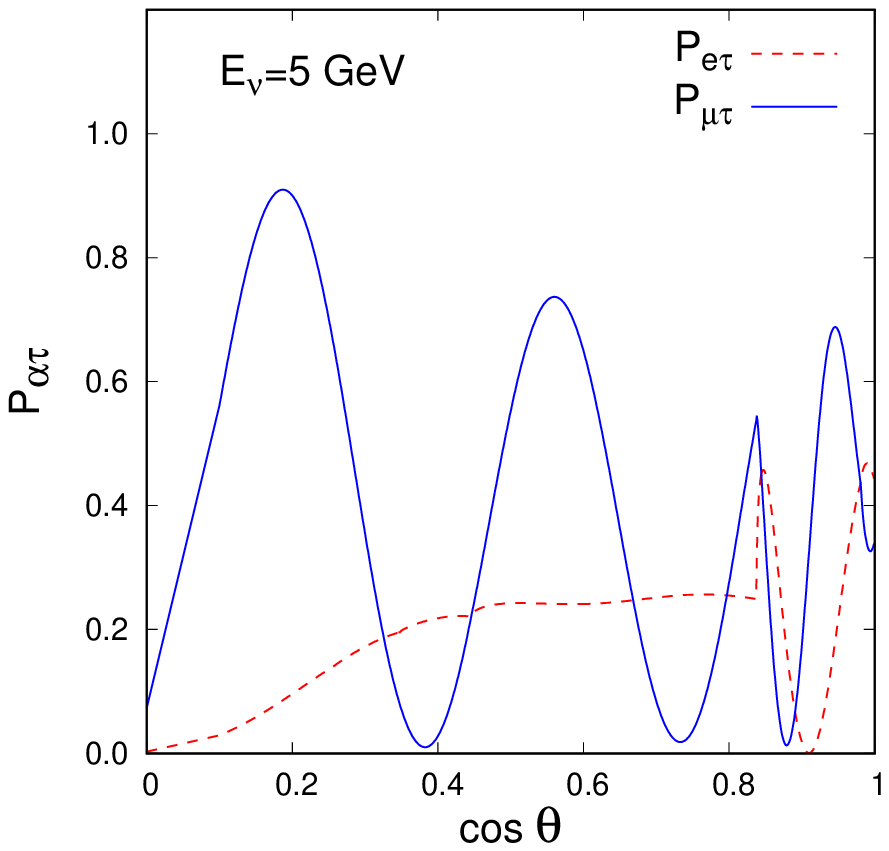}
\caption{~}
         \label{fig:P5}
     \end{subfigure}
     \hfill
     \begin{subfigure}[t]{0.49\textwidth}
         \centering
         \includegraphics[width=\textwidth]{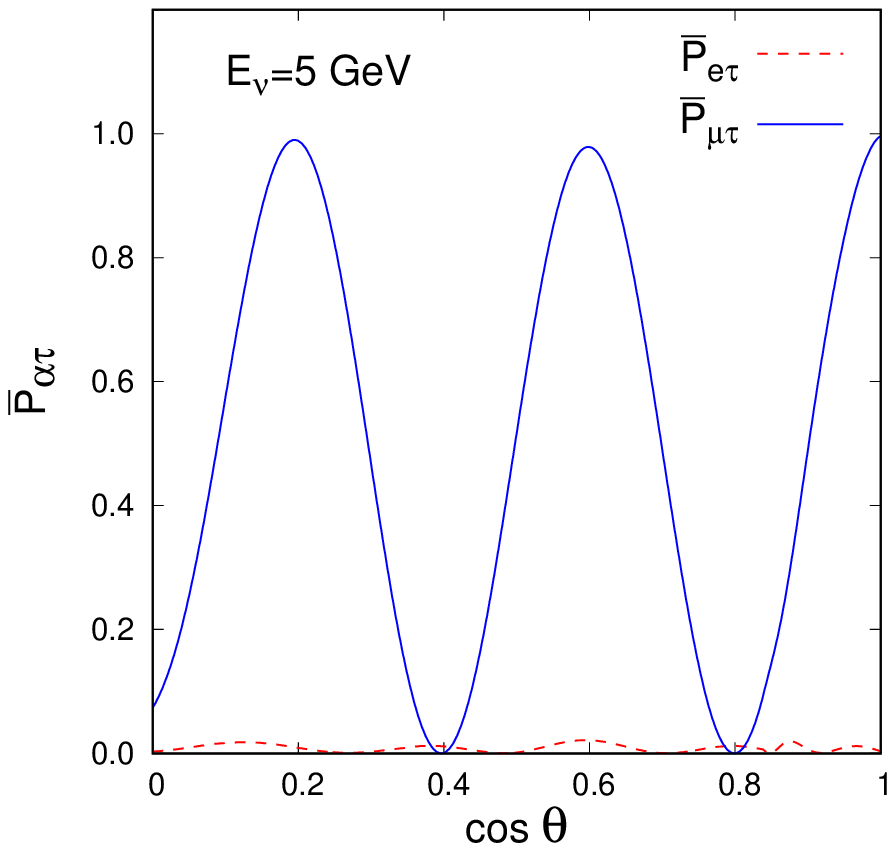}
\caption{~}
         \label{fig:P5bar}
     \end{subfigure}
     \begin{subfigure}[t]{0.49\textwidth}
         \centering
         \includegraphics[width=\textwidth]{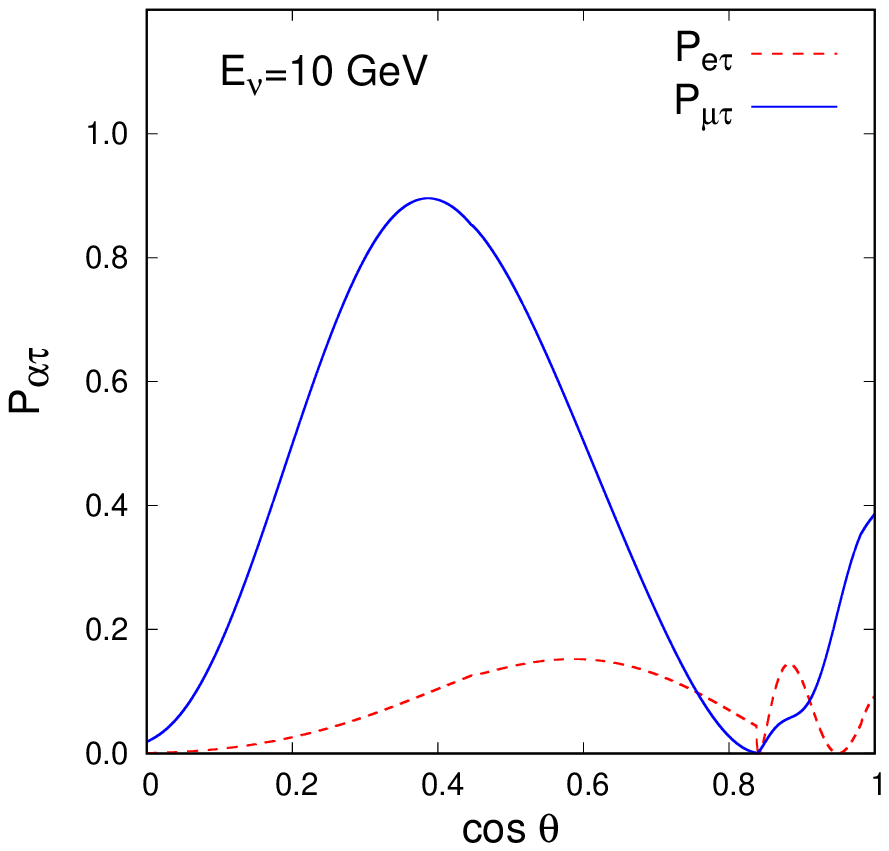}
\caption{~}
         \label{fig:P10}
     \end{subfigure}
     \hfill
     \begin{subfigure}[t]{0.49\textwidth}
         \centering
         \includegraphics[width=\textwidth]{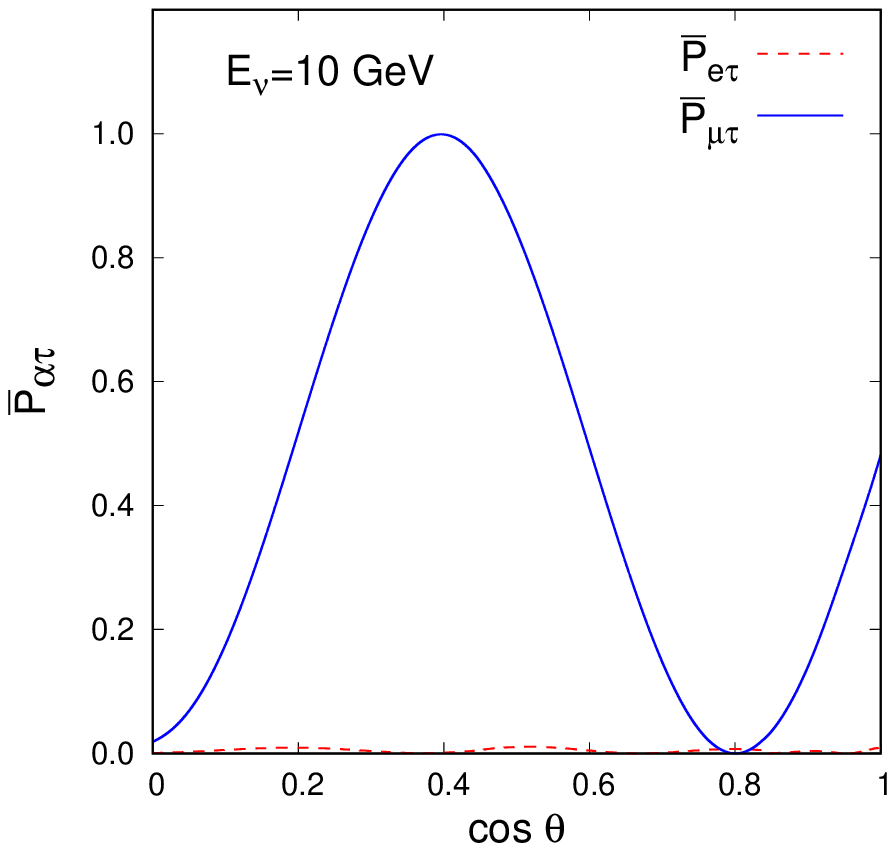}
\caption{~}
         \label{fig:P10bar}
     \end{subfigure}
\caption{Oscillation probabilities (a) $P_{\alpha\tau}$ for neutrinos
with $E_\nu=5$ GeV, (b) $\overline{P}_{\alpha\tau}$ for antineutrinos
with $E_\nu = 5$ GeV, $\alpha = e, \mu$,
as a function of the zenith angle $\cos\theta$
($\cos\theta = 1$ corresponds to UP neutrinos), with (c) and (d)
corresponding to the same plots for energy
$E_\nu = 10$ GeV respectively.}
\label{fig:Pat_E}
\end{figure}

We use these values of neutrino oscillation parameters to generate the
CC-tau events in ICAL. The threshold for this production is $E_{\rm th} =
3.5$ GeV. Since the atmospheric neutrino fluxes fall steeply with
energy, hence the dominant contributions to CC tau events are at
neutrino energies around 5--10 GeV. The tau oscillation probabilities were
computed using the Preliminary Reference Earth Model (PREM) profile
for the density distribution in the Earth using a 3-flavour model
\cite{Dziewonski:1981xy}. In the simulations, neutrinos of different
flavours were propagated through the atmosphere and through Earth matter
using a fast Runge-Kutta solver \cite{Indumathi:2006gr,Mohan:2016gxm}. The
oscillation probabilities $P_{\alpha\tau}$, $\alpha = e, \mu$, are shown
in Fig.~\ref{fig:Pat_E} as a function of the zenith angle $\cos\theta$
for a neutrino energy of $E_\nu = 5, 10$ GeV for both neutrinos and
anti-neutrinos, using the central values of the oscillation parameters
given in Table \ref{tab:osc}.

It can be seen that $P_{\mu\tau}$ is much larger than $P_{e\tau}$
so that the contribution from intrinsic muon atmospheric neutrinos
and anti-neutrinos will dominate over the contribution from electron
neutrinos.

Note also the discontinuity and features in both $P_{e\tau}$ and
$P_{\mu\tau}$ at $\cos\theta \simeq 0.86$, which corresponds to the
core-mantle boundary occurring in the neutrino oscillation probabilities
due to the choice of the normal mass ordering. Such features will instead
occur in $\overline{P}_{\alpha\tau}$ if the inverted ordering was assumed.

While horizontal events are hard to measure due to the geometry of ICAL,
it can be seen that the maximum contribution to tau events arises from the
region $\cos\theta \sim 0.5$ with a smaller contribution at $\cos\theta
\sim 1.0$ (due to the larger cross sections, neutrino events are about
three times larger than anti-neutrino events).

\begin{figure}[htp]
     \centering
     \begin{subfigure}[t]{0.49\textwidth}
         \centering
         \includegraphics[width=\textwidth]{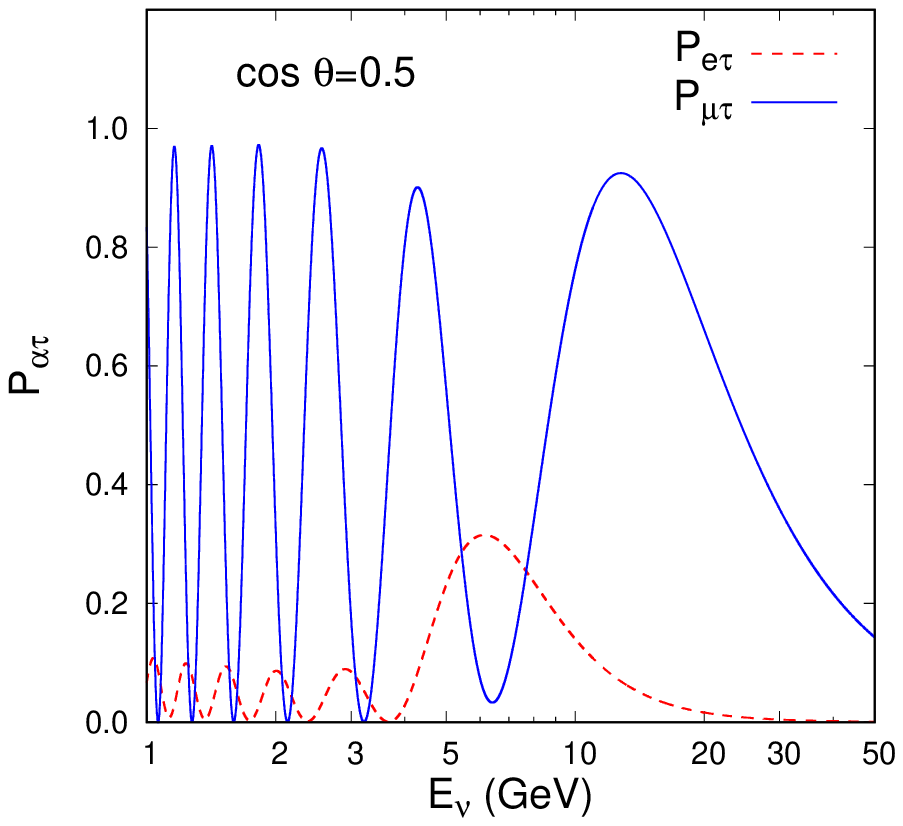}
\caption{ }
         \label{fig:Pcos}
     \end{subfigure}
     \hfill
     \begin{subfigure}[t]{0.49\textwidth}
         \centering
         \includegraphics[width=\textwidth]{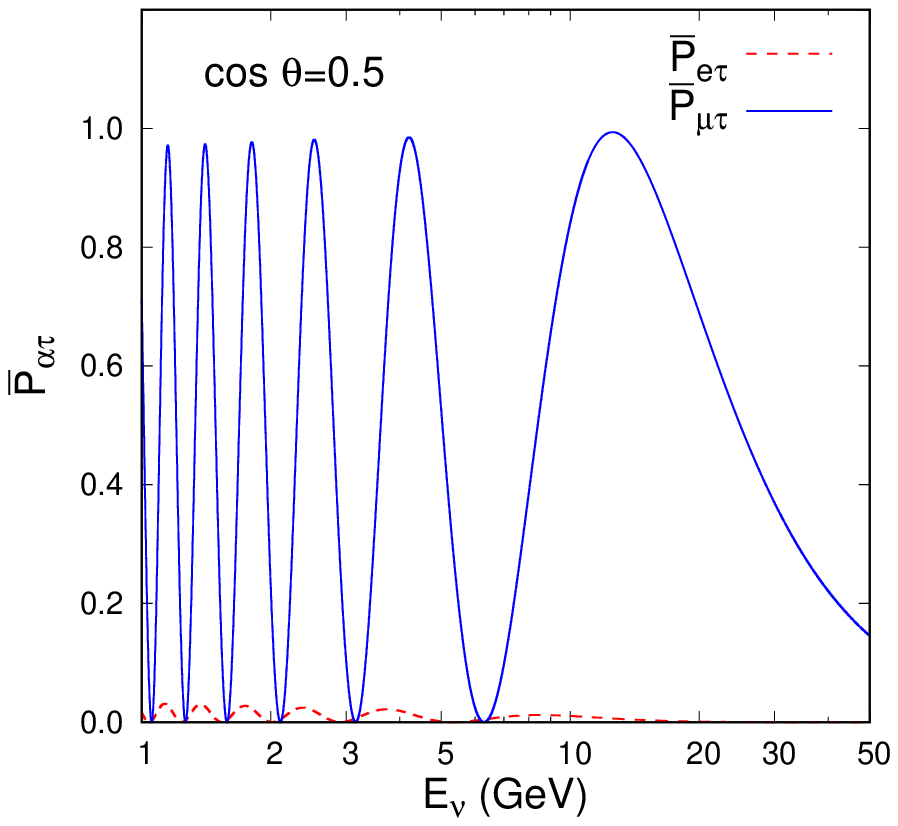}
\caption{ }
         \label{fig:Pbarcos}
     \end{subfigure}
\caption{Oscillation probabilities (a) $P_{\alpha\tau}$ for neutrinos
and (b) $\overline{P}_{\alpha\tau}$ for anti-neutrinos, $\alpha = e,
\mu$, as a function of the energy, $E_\nu$, for a zenith angle of
$\cos\theta =0.5$.}
\label{fig:Pat_cos}
\end{figure}

The tau oscillation probabilities $P_{\alpha\tau}$, $\alpha = e, \mu$,
are shown in Fig.~\ref{fig:Pat_cos} as a function of energy for a zenith
angle of $\cos\theta = 0.5$ ($\theta = 60^\circ$) for both neutrinos
and anti-neutrinos, again using the central values of the oscillation
parameters given in Table \ref{tab:osc}.

Note the presence of a broad maximum at $E_\nu \sim  10$--15 GeV. We
will see in the next section that this gives rise to CC tau events
with hadron energies of 5--10 GeV, to which ICAL has reasonably good
sensitivity. Before going on to the analysis of the sensitivity of CC-tau
+ NC events to the neutrino oscillation parameters, we briefly describe
the simulation of the ICAL detector and its response to the hadrons of
interest here.

\subsection{Hadrons in ICAL}
The proposed ICAL detector comprises 151 layers of 56 mm thick iron
plates placed 40 mm apart, interleaved with the resistive plate chambers
(RPCs) which are the active detector elements. This 51 kton detector will
be magnetised upto about 1.4 T although this does not affect the hadron
response \cite{Devi:2013wxa}. When charged particles pass through the RPCs,
they leave electrical signals called ``hits" that are picked up which
are localised to $3 \hbox{ cm} \times 3 \hbox{ cm} \times 0.2 \hbox{
cm}$ in the $x,y,z$ directions. The localisation in the vertical $z$
direction is due to the small 2 mm gas gap in the RPCs. While the
minimum ionising muons leave long tracks in the detector, the hadrons shower
and hence can be calibrated in energy and direction only from the shape
and number of hits in the shower \cite{Datta:2021myx}. A detailed study
of the hadron energy response from an analysis of these hits can be found
in Ref.~\cite{Devi:2013wxa} and has been used as-is in this analysis. A
brief study of the dependence of the sensitivity to neutrino oscillation
parameters on the hadron energy response is presented later in the
paper since the analysis relies heavily on this observable. Here we will
only highlight the angular response of the detector since it plays an
important role in the analysis, due to the upward nature of tau events.

As already mentioned, we are interested here in the events where the
tau decays hadronically. There are two sets of hadrons in such events:
the set of hadrons comprising the final state labelled $X$, which are
the hadrons produced in the original interaction, and those labelled $H'$
arising from tau decay.

\begin{figure}[btp]
\centering
\includegraphics[width=0.5\textwidth]{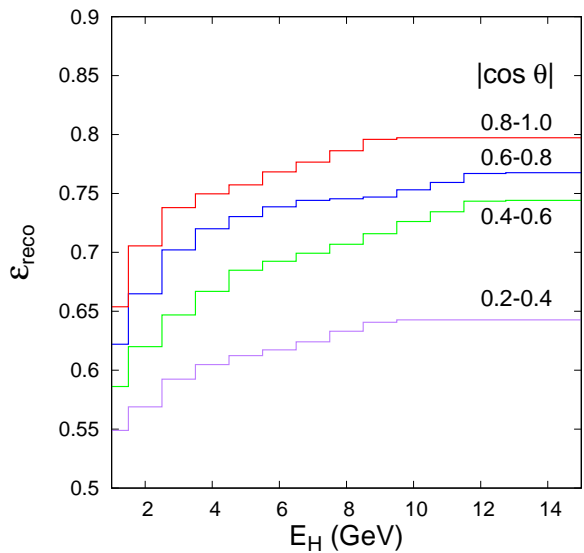}
\caption{Hadron angular reconstruction efficiency $\epsilon_{reco}$
as a function of the total hadron energy, for $\vert \cos\theta
\vert$ in the ranges 0.2--0.4, 0.4--0.6, 0.6--0.8,  and 0.8--1.0 from
an analysis of data simulated in Ref.~\cite{Devi:2018ltf}; it can be
seen that $\epsilon_{reco}$ increases as $\cos\theta$ increases.}
\label{fig:anglereco}
\end{figure}

Since the sets of hadrons cannot be distinguished and the tau decays rapidly
($\tau_\tau = 2.9\times 10^{-13}$ s), all hadrons in each event are
detected as a single shower in the detector. Due to the kinematics of
both tau production and decay, the hadrons are peaked in the direction
of the incident neutrino \cite{Indumathi:2009hg}. Earlier simulations studies
of hadrons arising from NUANCE events showed that the
hadron shower can be identified and their total energy calibrated by the
hits in the RPCs, when the hadrons pass through them
\cite{Devi:2013wxa}. In addition, the
direction of these hadrons can also be determined, although quite crudely
compared to the direction of muons \cite{Devi:2018ltf}. Due to the
geometry of ICAL (with horizontal iron plates), the reconstruction
ability in the vertical direction is expected to be better than for
more horizontal events. While it was shown that the zenith
angle in vertical directions can be reconstructed to within $10^\circ$
for 10 GeV hadrons, we are here interested only in the efficiency with
which an upward/downward going hadron is reconstructed as an
upward/downward going hadron, that is, reconstructed in the correct
quadrant, and not in the actual zenith angle itself. The result of the
simulation study is shown in Fig.~\ref{fig:anglereco} where the angular
reconstruction efficiency of hadrons is shown as a function of both the
true hadron energy and angle ($\cos\theta$). Here the reconstruction
efficiency is defined as
\begin{equation}
\epsilon_{\rm reco} = \frac{\hbox{Number of events reconstructed in the
correct quadrant}}{\hbox{Total number of events reconstructed}}~.
\end{equation}
It is seen that the best direction reconstruction occurs at higher
energies and for vertical angles (as expected), and is just about greater
than 50\% at more horizontal angles.  We use this information in the
analysis to separate the $\tau$ events into two direction bins, viz.,
up and down.

The neutrino energy and angle are used for calculating the oscillation
probabilities, while the parameters used in the analysis are the
summed reconstructed hadron energy in the event (denoted as $E_H = E_X
+ E_H'$, where $E_X = E_\nu - E_\tau$, where $E_\tau$ is the final lepton
energy), and the direction of the hadron shower (in two bins of UP/DOWN
alone). From the direction reconstruction efficiency plot shown in
Fig.~\ref{fig:anglereco}, we see that a fraction $(1-\epsilon_{reco})$
of UP events will be reconstructed as DOWN events and vice versa. Details
of the hadron energy resolution used in this analysis can be found in
Ref.~\cite{Devi:2013wxa}.

\subsection{Generation of oscillated events at ICAL}

The hadron energy and angle in the oscillated CC-tau events are smeared
into the observed hadron energy and angle event-by-event, as per the
detector response described above, and binned appropriately. Similarly,
the NC events generated from NUANCE are also binned in the same bins.
The dependence on the obtained sensitivity to the neutrino oscillation
parameters due to errors in the hadron energy response is discussed in
Section~\ref{ssec:eh}.

\begin{figure}[htp]
     \centering
     \begin{subfigure}[t]{0.49\textwidth}
         \centering
         \includegraphics[width=\textwidth,height=0.9\textwidth]{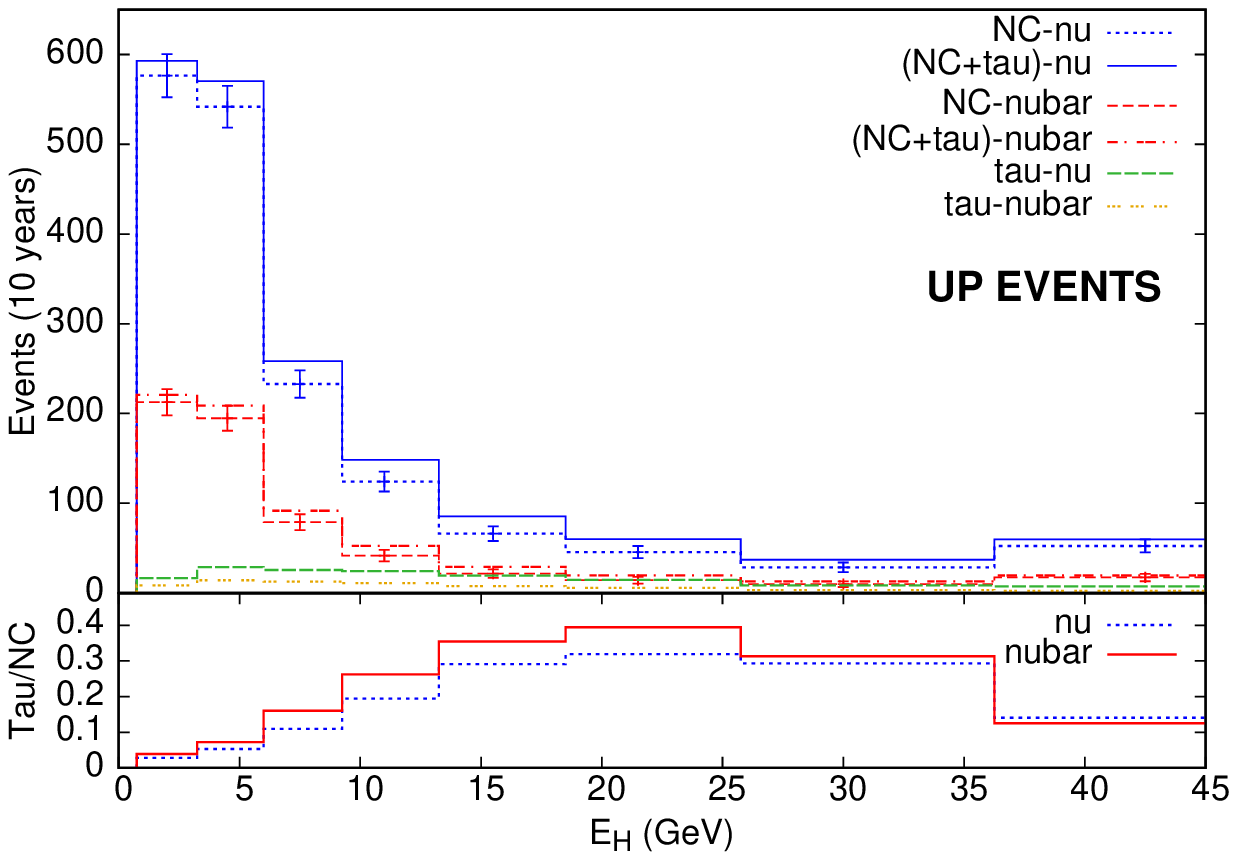}
\caption{ }
         \label{fig:up}
     \end{subfigure}
     \hfill
     \begin{subfigure}[t]{0.49\textwidth}
         \centering
         \includegraphics[width=\textwidth,height=0.9\textwidth]{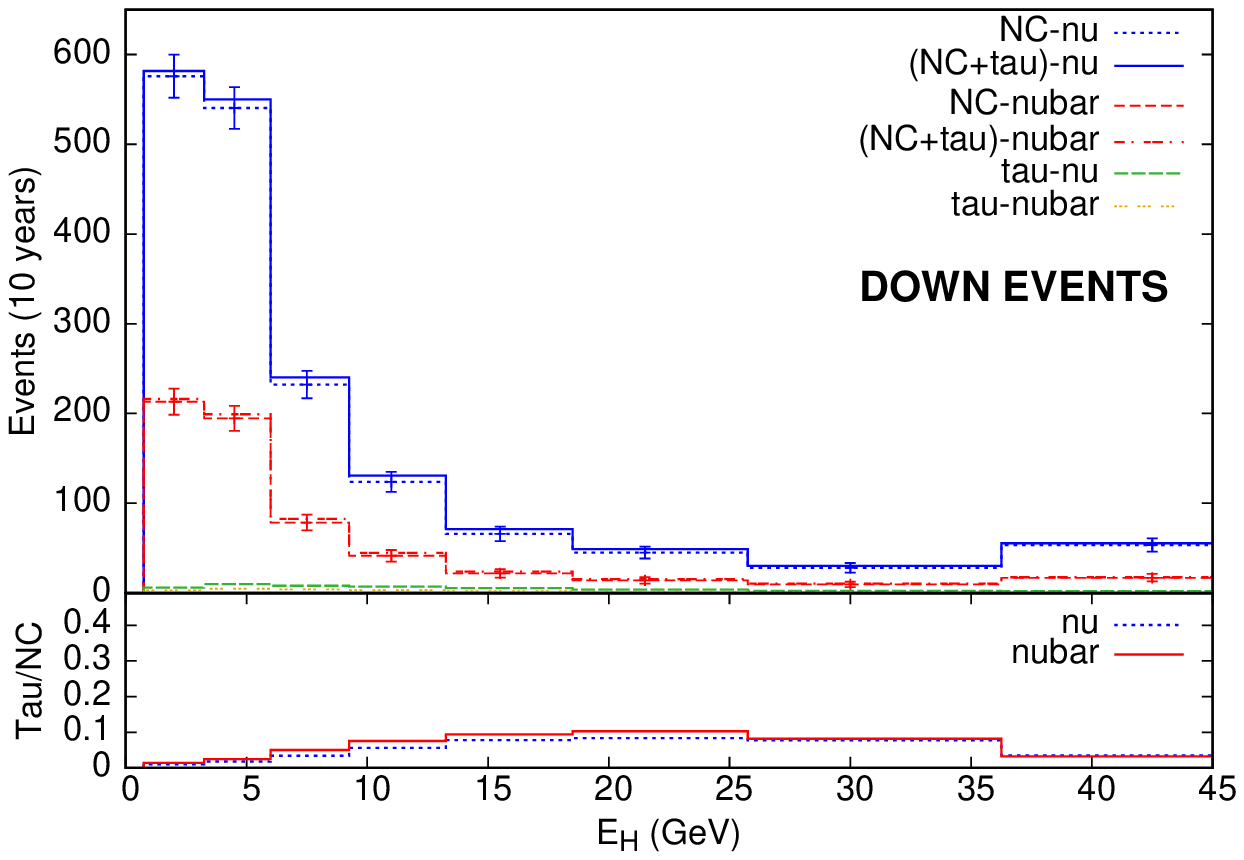}
\caption{ }
         \label{fig:down}
     \end{subfigure}
\caption{10 year sample of NC and CC-tau events in bins of
reconstructed hadron energy $E_H$ for (a) upward going (UP) events and (b)
downward going (DOWN) events. The contributions from NC and (CC-tau + NC)
events arising from neutrino (nu) and anti-neutrino (nubar) interactions
in the detector and the individual contributions from CC tau events
are shown. The panel below shows the ratio of the CC tau events to the
NC events.}
\label{fig:ncandcc}
\end{figure}

Fig.~\ref{fig:ncandcc} shows the energy distribution of a 10 year sample
of CC-tau and NC events separately from neutrino and antineutrino sources
(although they are not separable in the detector). The CC-tau events have
been generated using the central values of the oscillation parameters
given in Table \ref{tab:osc}. Fig.~\ref{fig:up} shows the events for which
the hadron shower direction is reconstructed as being in the UP direction
($\cos\theta > 0$) while Fig.~\ref{fig:down} is for the DOWN events
($\cos\theta < 0$). It can be seen that there are a small number of
CC-tau events reconstructed in the DOWN bin due to the error in direction
reconstruction, as seen already from Fig.~\ref{fig:anglereco}. However,
the increase over the NC background is clearly visible for events in the
UP bin, where the error bars shown are statistical. We are now ready
to analyse these events for their sensitivity to neutrino oscillation
parameters.

\section{Numerical analysis and physics reach}

We now go on to a numerical study to examine the sensitivity of NC + CC
tau events to various neutrino oscillation parameters of interest.

\subsection{Generation of data samples}

Statistical fluctuations are significant in the analysis of neutrino
events for their sensitivity to the oscillation parameters. This is
in fact what distinguishes the sensitivity for different exposure
times. The statistical limitations of data contribute in two differing
ways. One is the error or precision with which a neutrino oscillation
parameter (say $\theta_{23}$) can be measured. It is determined by the
amount of data available for the analysis. The other limitation is the
fluctuations present in the data sample which may yield a best
fit value for the parameter which may not coincide with its true value.
In simulations analysis, therefore, different
sets of ``data" samples, all randomly generated for the same $n$ number
of years, will
yield different best values. Hence using a single arbitrary ``data"
sample in such simulations analysis
raises the risk of over- or under-estimation of sensitivity for a
given input value. This can be avoided by repeating the analysis
$N$ times with different ``data" sets of generated data samples for the given
input value. The average sensitivity in these repeated analyses (for
$N \gtrsim 60$), will approach a median sensitivity for the given input
value of oscillation parameter; in fact, the various results obtained
with each sample will cluster around this median sensitivity.

Alternatively, using a procedure which enormously saves computational
time, a large data sample (1000 years) is generated and scaled to
the required $n$ number of years (typically 10 years) during the analysis.
This procedure will yield the median sensitivity for a given input
value of parameters in the $\chi^2$ analysis for which the sensitivity
is being measured. It was shown in several analyses
\cite{Blennow:2013oma,Cowan:2010js}
that such a procedure of generating large data samples and scaling them
to the required number of years correctly determines the precision with
which parameters can be determined.
We will use this technique for our analysis in what follows and examine
this in more detail with an example in Section~\ref{ssec:fluct}.

\subsection{Best fit approach}

A large data sample (1000 years) of CC-tau events was generated as per
the process described above. A similar 1000 year sample of NC events was
generated as well. Notice that the source of atmospheric neutrino fluxes
is the same in both cases. The simulated ``data'' is generated by applying
neutrino flavour oscillations using a set of input values of oscillation
parameters (typically the central values of the oscillation parameters
given in Table \ref{tab:osc}), and scaling this sample to 10 years, as
required. In order to test the sensitivity of this sample to neutrino
oscillations, the original unoscillated sample is oscillated using a
different value of one or more oscillation parameters, scaled to the
same number of years, and labelled ``theory''. The $\chi^2$ for this
set of ``data" and ``theory" is defined as
\begin{equation}
\chi^{2}  = \min_{\xi_{k}} \sum^{N_{E}}_{i=1} \sum^{N_{cos \theta}}_{j=1} \,
2 \left( \left( T^{ij} - D^{ij} \right) -
D^{ij} \ln \left( \frac{T^{ij}}{D^{ij}} \right) \right) +
\sum_{k=1}^{N_k} \xi_{k}^{2}~,
\label{eq:chisq}
\end{equation}
where,
\begin{description}
\item $D^{ij}= [D^{ij}_{\tau,+} + D^{ij}_{\tau,-} + D^{ij}_{NC,+} +
D^{ij}_{NC,-}]$ is the total number of CC-tau and NC events arising from
both antineutrino ($+$) and neutrino ($-$) fluxes, generated with the
set of input values of oscillation parameters, in the $i^{th}$ energy
and $j^{th}$ $\cos \theta$ bin, defined to be the ``Data" in the
simulations study,

\item $T^{ij}$ is the corresponding number of predicted theory events
in the same bins, generated using a different set of oscillation
parameters, where we have included the systematic uncertainties via
the pulls technique; see Refs.~\cite{Kameda,Ishitsuka}, so that the number
of theory events includes the systematic uncertainty from five sources
which are described in detail in Section \ref{ssec:syst}:
\begin{eqnarray} \nonumber
T^{ij} & = &  T^{ij}_+ + T^{ij}_-~, \\
\hbox{where   } T^{ij}_\pm & = &  T^{ij,0}_\pm
	\left( 1 + \sum_{k=1}^{N_k} \pi^{ij}_{k,\pm} \xi_{k,\pm} \right)~.
\label{eq:th}
\end{eqnarray}
Here $T^{ij,0}_\pm$ are the number of antineutrino/neutrino theory
events, without systematic errors in the corresponding bins, and

\item $\xi_k^2 \equiv \xi_{k,-}^2 + \xi_{k,+}^2$ includes the penalty
from each pull parameter for neutrino and anti-neutrino events
respectively.
\end{description}
A measure of the sensitivity of the data to the input value of any
parameter is given by $\Delta \chi^2$, defined as
\begin{equation}
\Delta \chi^2 = \chi^2(\hbox{par}) - \chi^2(\hbox{input})~,
\label{eq:delchi2}
\end{equation}
where $\chi^2(\hbox{input})$ corresponds to the minimum $\chi^2$
obtained when the theory events are calculated with the same value of
the parameter as its input value, and $\chi^2(\hbox{par})$ is the
minimum value obtained when a different theory value of the parameter
is used. Note that when the scaling procedure is used to generate the
``data", the value of $\chi^2(\hbox{input})=0$. The definition can be
appropriately extended to the case when more than one parameter is varied
from its input value when calculating the ``theory" events.

Finally, a prior on the well-known parameter $\sin^22\theta_{13}$ is
included via
\begin{equation}
\chi^{2}_{tot} = \chi^{2} + \left( \frac{\sin^{2}2 \theta_{13}^{in} -
\sin^{2}2 \theta_{13} }{\sigma_{\sin^2 2 \theta_{13}}} \right)^{2}~.
\label{eq:prior}
\end{equation}
Note that $\sin^22\theta_{13}$ is very well constrained; also, while
the sensitivity of ICAL to the neutrino mass ordering depends on this
parameter, it is not very sensitive to the value of this parameter
itself, and hence any changes in the central values used for this
parameter will not affect our results. While including the systematics,
the minimisation with respect to the systematic nuisance parameters is
done by analytically solving for the set of $\xi_i$ for a given set of
oscillation parameters. The minimum $\chi^2$ is then found by varying the
oscillation parameters for the ``theory" set of events and marginalising
over irrelevant parameters as required.

\subsection{Systematic uncertainties in the analysis}
\label{ssec:syst}
The inherent systematic uncertainties associated with the prediction of
the events and their rates affect the sensitivity of these events to the
oscillation parameters. There are five different types of systematic
unertainties which are considered in our analysis,
{\em viz}., the relevant sum in Eqs.~\ref{eq:chisq} and \ref{eq:th}
runs over $k = 1, \cdots, N_k=5$ \cite{ICAL:2015stm}. These values are
standardly used by the INO Collaboration in all its analyses
\cite{ICAL:2015stm}. 

\begin{enumerate}
\item We calculate the energy dependent flux tilt error by considering a
deviation of $\delta =5$\% from the standard behaviour, viz.,
$E_\nu^{-2.7}$. Hence the systematic error $\pi_{tilt}$ from this uncertainty
can be calculated for each energy bin through
\begin{eqnarray} \nonumber
\Phi_{\delta}(E) &= & \Phi_{0} (E)\left(
\frac{E}{E_{0}}\right)^{\delta}~,  \\
 &\approx & \Phi_{0} (E) \left( 1 + \delta ~ \ln \frac{E}{E_{0}}
 \right)~.
\end{eqnarray}
We choose the standard value of $E_{0} = 2$ GeV \cite{Honda:2011nf}.

\item We consider the flux angular uncertainty to be
$\pi_{zenith}=5$\% $\cos\theta$, in the given zenith angle bin.

\item We take the overall flux normalisation uncertainty to be
$\pi_{norm} = 20$\%.

\item The systematic error due to uncertainty in computing the cross
section is taken to be $\pi_\sigma = 10$\%.

\item Finally, an overall uncertainty of $\pi_D=5$\% is included to take
care of any uncertainty in characterising the detector response.

\end{enumerate}
We have summarised the systematic uncertainties in Table \ref{tab:sys}.
Note that we have taken the uncertainties to be the same for the neutrino
and antineutrino samples, and these events are combined to compute the
$\chi^2$ since they are not distinguishable in the detector. Note also
that the last three overall bin-independent errors can be replaced
numerically by a single one: $\pi = \sqrt{\pi_{norm}^2 + \pi_\sigma^2 +
\pi_D^2}$; they are separately retained to allow for later refinement
in the inclusion of systematic errors. We discuss in Section~\ref{ssec:eh}
the effect of an energy-dependent uncertainty while implementing the
hadron energy response of the detector, which is a preliminary study on
the bin dependence of the pull corresponding to the detector response,
$\pi_D$.

\begin{table}[htp]
\centering
\vspace{0.2cm}
\begin{tabular}{| l | c | r |}
\hline
Pull ($\pi$) & Description & Error \\
\hline
$\pi_{tilt}$ & Tilt error & 5\% \\
$\pi_{zenith}$ & Zenith angle dependence & 5\% \\
$\pi_{norm}$ & Flux normalization & 20\% \\
$\pi_{\sigma}$ & Cross section & 10\% \\
$\pi_{D}$ & Detector response & 5\% \\
\hline
\end{tabular}
\caption{Systematic uncertainties included in the analysis.}
\label{tab:sys}
\end{table}

\subsection{The binning scheme}
We have optimised the number of hadron energy bins by computing
the $\chi^2$ values for various sets of data and theory. The final set of
bins used in the analysis is listed in Table \ref{tab:hadbin}.

\begin{table}[htp]
\begin{center}
\begin{tabular}{|c|c|c|} \hline
Bin & Energy range & Bin width \\
    &  (GeV) & (GeV) \\ \hline
1 & 1--3 & 2 \\
2 & 3--6 & 3 \\
3 & 6--9 & 3 \\
4 & 9--13& 4 \\
5 &13--18& 5 \\
6 &18--25& 7 \\
7 &25--35& 10 \\
8 &35--50 & 15 \\ \hline
\end{tabular}
\end{center}
\caption{Hadron energy bins used in the analysis.}
\label{tab:hadbin}
\end{table}

\subsection{Sensitivity to the presence of tau events}

We begin by asking whether ICAL will be sensitive to the presence of
the CC-tau events in the sample. That is, we consider the situation
where the ``theory" does not account for the presence of tau neutrinos
in the sample. Hence the theory events arise purely from the NC events
and are independent of neutrino oscillations. We find that, without
including sytematic uncertainties, the significance to the presence of
tau is at a confidence level of about $\sqrt{\Delta \chi^2} \sim 6\sigma$
for 10 years exposure; when the systematic uncertainties are included as
described above, this decreases to $3.6\sigma$; see Table~\ref{tab:tau}.
The major impact is due to the tilt and the zenith angle uncertainties,
while the impact of the remaining uncertainties is small or negligible.
For instance, if the zenith angle and tilt uncertainties reduce to 70\%
or 50\% of the values given in Table~\ref{tab:sys} by the time ICAL is
operational, the sensitivity will increase to $3.9\sigma$ and $4.1\sigma$
respectively. Hence it is possible that sensitivity to the presence of
tau neutrinos can be achieved to nearly $4\sigma$ level with ICAL.

\begin{table}[htp]
\centering
\begin{tabular}{|c|c|c|c|} \hline
Angular Bins & Incl. Systematics & $\Delta \chi^2$ &
Significance = $\sqrt{\Delta \chi^2}$ \\ \hline
2 & No & 35.34 & 5.95 \\
2 & Yes & 13.06 & 3.61 \\
2 & Yes$^*$ (0.7) & 14.95 & 3.87 \\
2 & Yes$^*$ (0.5) & 16.46 & 4.06 \\ \hline
1 & No & 27.50 & 5.24 \\
1 & Yes & 7.90 & 2.81 \\ \hline
\multicolumn{4}{|l|}{*See text/caption for details} \\ \hline
\end{tabular}
\caption{Sensitivity to the presence of tau neutrino events in various
scenarios including the number of angular bins, and whether systematic
uncertainties were included or not. The effects of a reduction in tilt and
zenith angle uncertainties to 70\% and 50\% of their values given in
Table~\ref{tab:sys} are also shown.}
\label{tab:tau}
\end{table}

Also note the effect of separating the events in UP/DOWN angle bins. If
no angular information is taken into account, the significance falls to
just $2.8\sigma$. This highlights the importance of angular information
in this analysis, due to the fact that almost all tau neutrinos are
produced in the upward direction. An improvement in the angular
reconstruction of hadron showers will therefore vastly improve this
result, which is comparable to that obtained by the SuperKamiokande
Collaboration also with atmospheric neutrino data
\cite{Super-Kamiokande:2017edb}. This is discussed further in the next
two sections.

\subsection{Sensitivity of tau neutrino events to the oscillation parameters}

We now proceed to study the sensitivity of the tau events (with NC
background) to the neutrino oscillation parameters, in particular,
to the 2--3 parameters $\sin^2\theta_{23}$ and $\Delta m^2$. We begin
with the mixing angle, $\sin^2\theta_{23}$. Now the data and theory
include both the NC and CC-tau events, with the ``theory'' events being
generated with a different value of $\sin^2\theta_{23}$ than the input
value of $\sin^2\theta_{23}^{in}$ used to generate the data. The resulting
$\Delta \chi^2$, defined in Eq.~\ref{eq:delchi2}, is then a measure of
the sensitivity of the data to the input value of the parameter.

\begin{figure}[thp]
     \centering
     \begin{subfigure}[t]{0.49\textwidth}
         \centering
         \includegraphics[width=\textwidth]{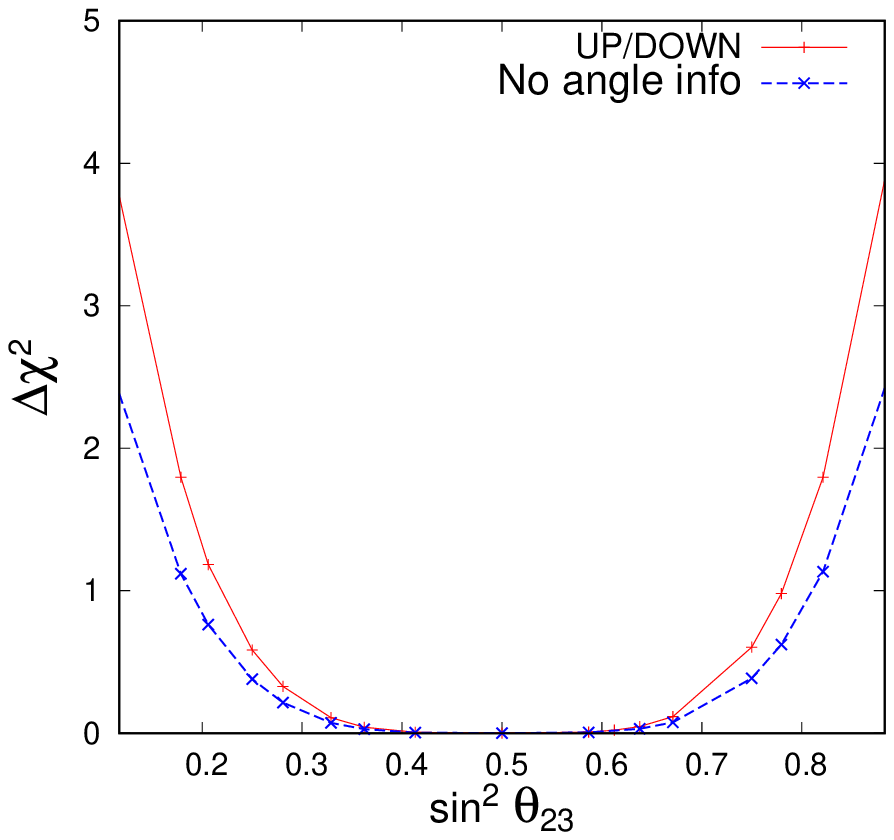}
\caption{ }
         \label{fig:t23marg}
     \end{subfigure}
     \hfill
     \begin{subfigure}[t]{0.49\textwidth}
         \centering
         \includegraphics[width=\textwidth]{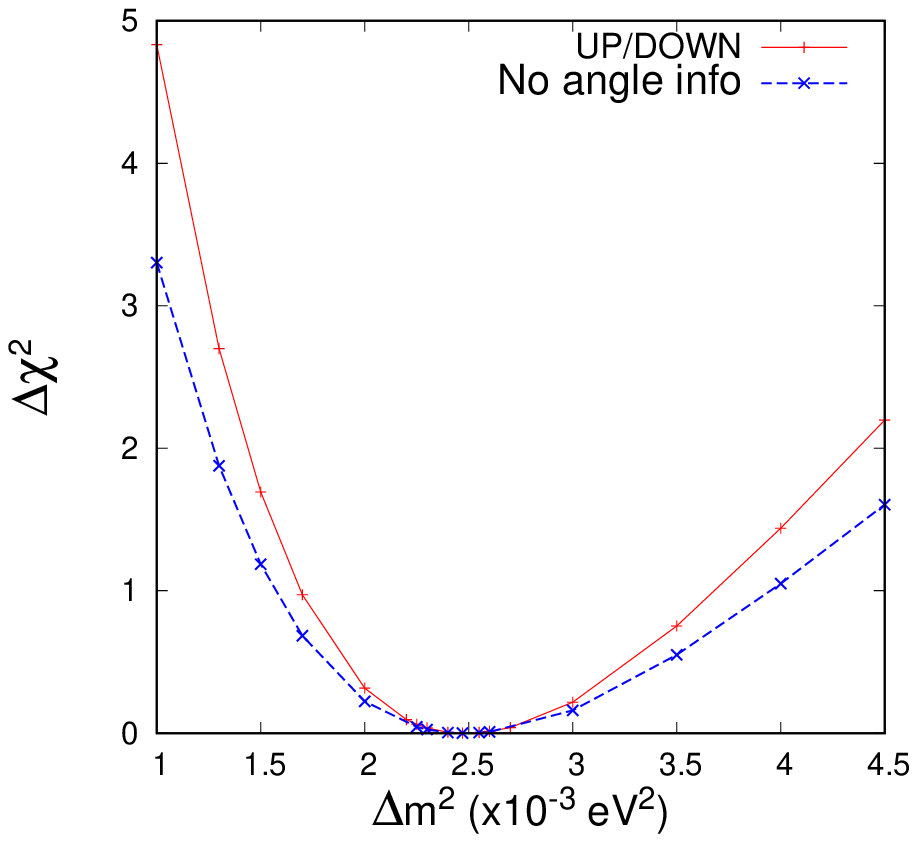}
\caption{ }
         \label{fig:dm2marg}
     \end{subfigure}
\caption{Sensitivity of CC-tau + NC events to the oscillation
parameters, $\sin^2\theta_{23}$ and $\Delta m^2$. Shown in (a) is
the change in $\chi^2$, $\Delta \chi^2$, when the theory value of
$\sin^2\theta_{23}$ is varied, for an input value of
$\sin^2\theta_{23}^{in}=0.5$ ($\theta_{23}^{in}=45^\circ$) while (b)
shows $\Delta \chi^2$ when the theory value of $\Delta m^2$ is varied,
for an input value of $\Delta m^2_{in} = 2.47 \times 10^{-3}$ eV$^2$.
The improvement on including two zenith angle bins, viz., UP and DOWN,
is also shown. For more details, see the text.}
\label{fig:t23zenith}
\end{figure}

Fig.~\ref{fig:t23marg} shows the resulting $\Delta \chi^2$ as a function
of the value of $\sin^2\theta_{23}$ used to generate the theory events,
for an input value of $\sin^2\theta_{23} = 0.5$ ($\theta_{23}^{in} =
45^\circ$). Here $\sin^2\theta_{23}$ is kept fixed while $\Delta m^2$
and $\sin^22\theta_{13}$ are marginalised over their $3\sigma$ ranges as
listed in Table \ref{tab:osc}; the systematic uncertainties mentioned
above are also included in the analysis. It can be seen that including
the zenith angle dependence (using the information on UP/DOWN bins shown
in Fig.~\ref{fig:anglereco}) improves the sensitivity even though the
hadron direction is poorly determined. The same trend is seen for
$\Delta m^2$, as seen in Fig.~\ref{fig:dm2marg}. Henceforth we
always consider the data binned in both hadron energy and angle.

It may be argued that including even more zenith angle bins will further
improve the sensitivity. Indeed, when the up-going events are binned
in 2 zenith angle bins ($0^\circ < \theta < 45^\circ$ and $45^\circ <
\theta < 90^\circ$) while retaining the down-going events in a single
bin, there is a small improvement in the sensitivity. However, due to
the limited reconstruction capability of the zenith angle of hadrons,
there will be large correlations between these bins; such an analysis
requires a deeper study of the angular reconstruction of hadrons in
ICAL than is available at present. In addition, there is also a modest
improvement when the hadron energy resolution is improved, for instance,
on using the energy resolution that would be obtained if ICAL used 2 cm
iron plates rather than the design value of 5.6 cm.

\begin{figure}[thp]
     \centering
     \begin{subfigure}[t]{0.49\textwidth}
         \centering
         \includegraphics[width=\textwidth, height=0.8\textwidth]{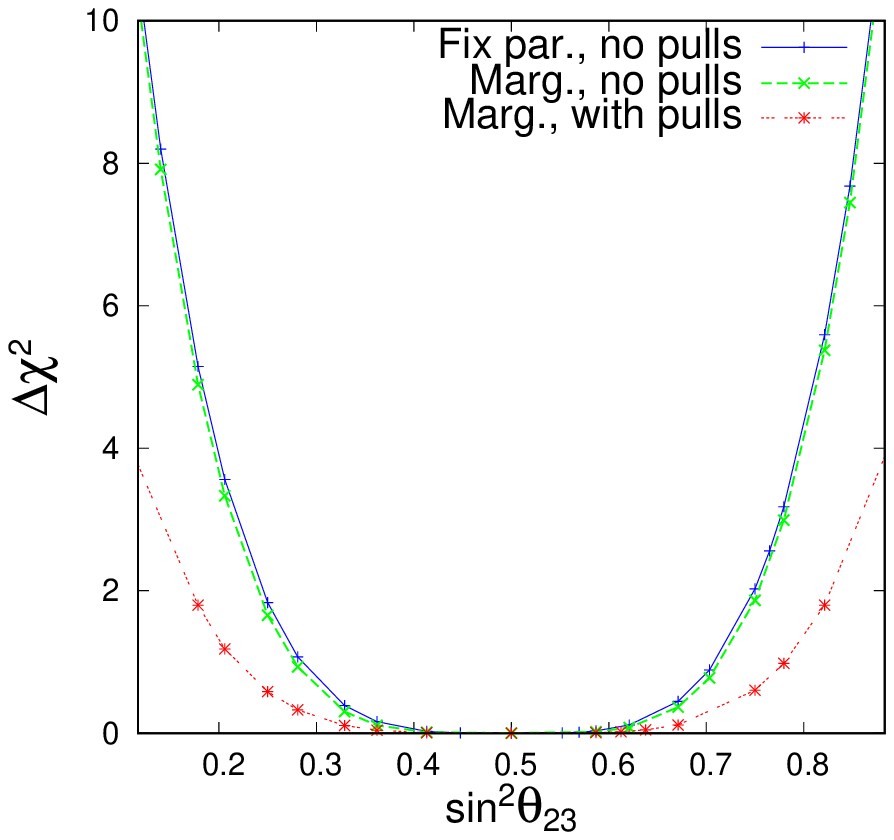}
\caption{ }
         \label{fig:t23only}
     \end{subfigure}
     \hfill
     \begin{subfigure}[t]{0.49\textwidth}
         \centering
	\includegraphics[width=\textwidth, height=0.8\textwidth]{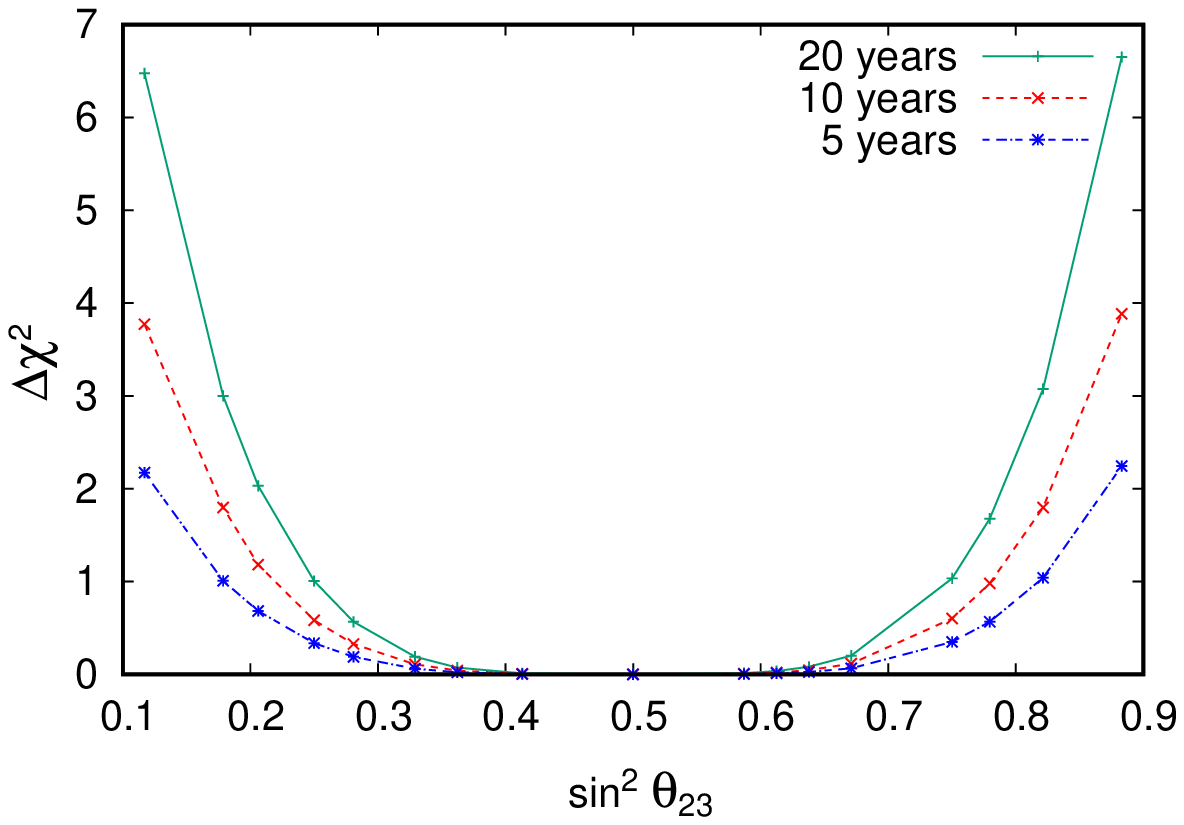}
\caption{ }
\label{fig:t23_yr}
     \end{subfigure}
\caption{L: $\Delta \chi^2$ as a function of the theory value of
$\sin^2\theta_{23}$, showing sequentially the effect of marginalisation
(of the dominant parameters $\Delta m^2$ and $\sin^22\theta_{13}$
over their $3\sigma$ ranges) and the effect of including systematic
errors, for input values $\sin^2\theta_{23}^{in} = 0.5$ and
$\Delta m^2_{in} = 2.47 \times 10^{-3}$ eV$^2$. R: 
$\Delta \chi^2$ as a function of exposure time of 5, 10 and 20 years
with systematics and marginalisation.}
\end{figure}

We find that the sensitivity is dominated by the systematic
uncertainties as can be seen in Fig.~\ref{fig:t23only} where the results
of an analysis with no consideration of systematic uncertainties and
either keeping all parameters fixed at their central values (other
than $\sin^2\theta_{23}$, of course) or marginalised over their $3\sigma$
ranges listed in Table \ref{tab:osc} are shown.

The analysis is insensitive to the value
of the CP phase and this as well as the 1--2 oscillation parameters have
been kept fixed to their central values listed in Table \ref{tab:osc}. In
addition, the neutrino ordering is assumed to be normal, unless stated
otherwise. While there is practically no effect on including
marginalisation, the curve labelled ``Marg., with pulls" indicates
clearly that the maximum loss of sensitivity occurs on including
systematic errors.

In Fig.~\ref{fig:t23_yr}, we see the sensitivity to the oscillation
parameter $\sin^2\theta_{23}$ as a function of the exposure time.  The
results correspond to inclusion of systematic errors and marginalisation
over the remaining parameters.  While the sensitivity to the oscillation
parameter precision measurement improves as the number of years increases,
this is not exactly linear in the number of years due to systematic
effects.

\begin{figure}[thp]
\centering
\includegraphics[width=0.60\textwidth]{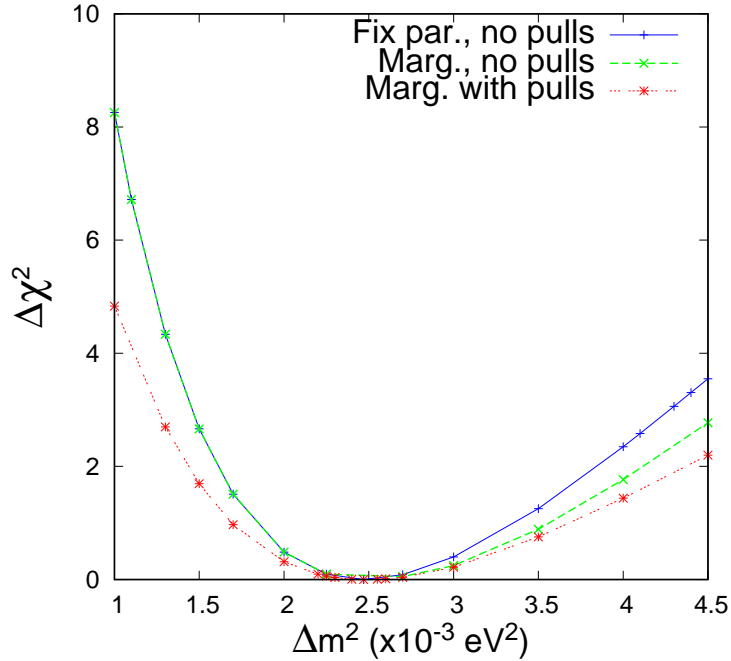}
\caption{$\Delta \chi^2$ as a function of the theory value of $\vert \Delta
m^2\vert$, for input value of $\Delta m^2_{in} = 2.47 \times 10^{-3}$
eV$^2$; here all other parameters are marginalised over their $3\sigma$
ranges.}
\label{fig:dm2only}
\end{figure}

In Fig.~\ref{fig:dm2only}, we have shown the
sensitivity study to the oscillation parameter $\vert \Delta m^2\vert$. It
is seen that the tau events are able to discriminate better against $\vert
\Delta m^2\vert$ values that are lower than the input value since the
dependence on this parameter is via the dominant $P_{\mu\tau}$ oscillation
probability and arises as $\sin^2 [\Delta m^2 L/E_\nu]$. Finally,
we remark that the analysis indicates negligible sensitivity to the
neutrino mass ordering and is therefore insensitive to the {\em sign}
of $\Delta m^2$ (that is, to the sign of $\Delta m_{31}^2$ or $\Delta
m_{32}^2$). However, we believe that this is the first study to explore
the possibility of extracting neutrino oscillation information from a
study of taus produced in the interactions of atmospheric neutrinos.
It is clear that tau events, in spite of their small number in absolute
terms, have significant sensitivity to the oscillation parameters
$\sin^2\theta_{23}$ and $\Delta m^2$. In what follows, we only consider
the realistic case of two zenith angle bins, viz., UP and DOWN, with
the inclusion of systematic errors.

\subsection{A discussion on scaling the data sample}
\label{ssec:fluct}

Here, we show with an example how the procedure of generating the ``data"
sample using the 1000 year generated events scaled to the required number
of $n$ years yields the correct precision with which the parameters
are determined. In this example, we calculate the $\Delta\chi^2$
values for a theory value of $\sin^2\theta_{23}^{par}=0.25$ ($\theta_{23}
= 30^\circ$) using 10 different ``data" sets of 10 years data generated
{\it without scaling}. Here the other oscillation parameters are kept
fixed and systematic uncertainties have been ignored, for clarity.

\begin{figure}[thp]
     \centering
     \begin{subfigure}[t]{0.49\textwidth}
         \centering
         \includegraphics[width=\textwidth]{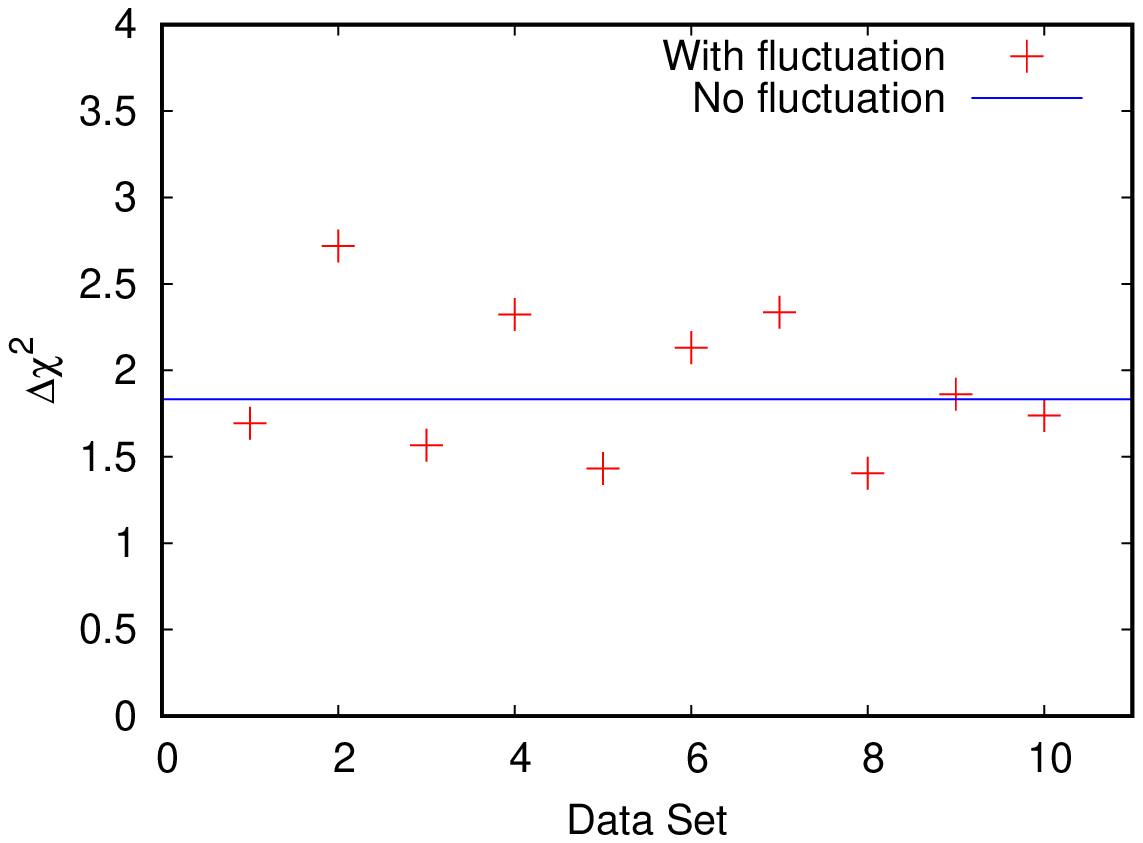}
\caption{ }
         \label{fig:fluct}
     \end{subfigure}
     \hfill
     \begin{subfigure}[t]{0.49\textwidth}
         \centering
         \includegraphics[width=\textwidth]{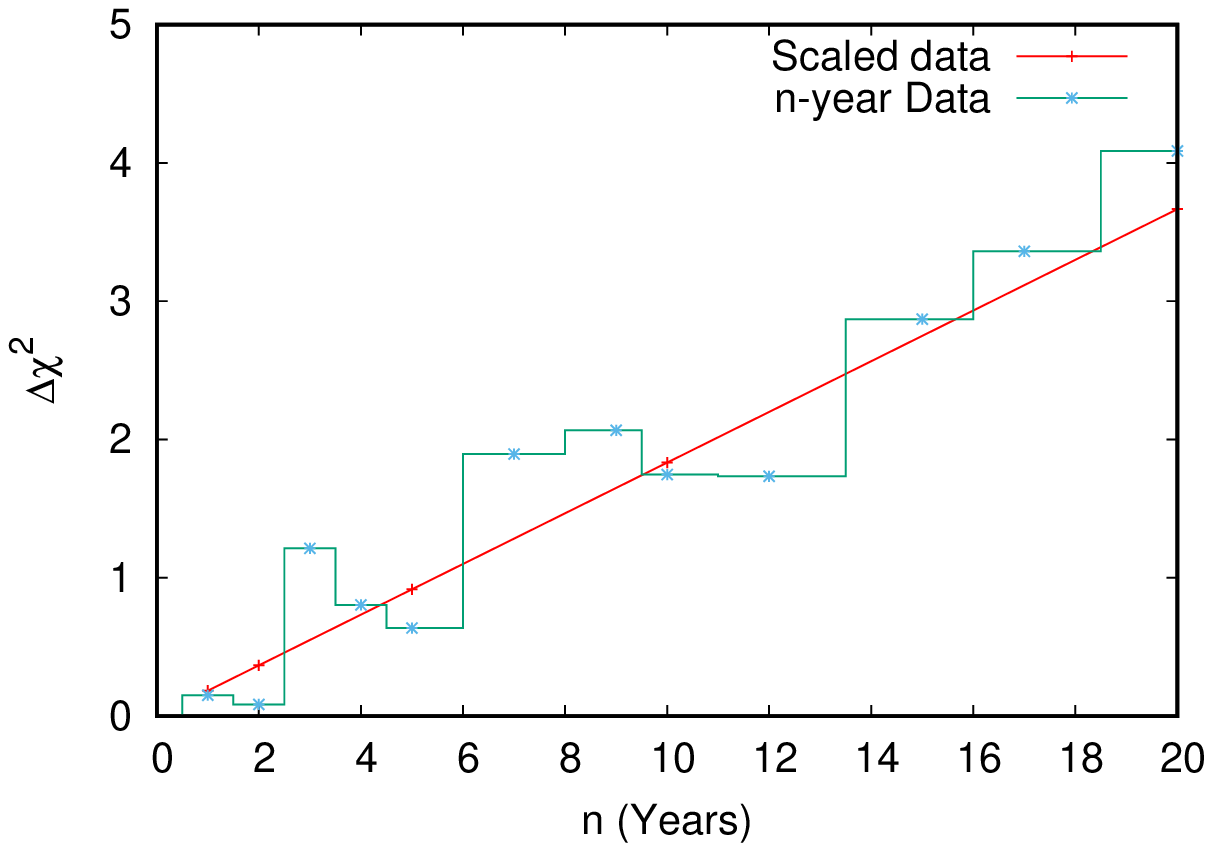}
\caption{ }
         \label{fig:fluctyear}
     \end{subfigure}
\caption{$\Delta\chi^2$ values obtained for
$\sin^2\theta_{23}^{par}=0.25$ ($\theta_{23} = 30^\circ$). Shown in
(a) is the result using 10 different samples of 10 years' data, labelled
1 through 10, with other parameters kept fixed. The solid line is the
result when the entire 1000 years events are scaled to 10 years and used
as ``data". Shown in (b) is the sensitivity with the two methods as a
function of different exposures in years at ICAL. The red smooth line
corresponds to the results with scaled data, while the green histogram
is obtained when one of the 10 data sets is used, with no scaling.}
\label{fig:t23dm2}
\end{figure}

As we see from Fig.~\ref{fig:fluct}, the calculated $\Delta\chi^2$
values for these samples are clustered around the value of $\Delta\chi^2
= 1.833$ which is the value obtained using the alternate method when
the 1000 year set is scaled to 10 years and used as ``data". It
clearly indicates the risk of over- or under-estimation of sensitivity
to $\sin^2\theta_{23}$ if we only use a single sample which was randomly
generated for 10 years.

We also further see in Fig.~\ref{fig:fluctyear} the consistency between
the procedure of scaling the events to the required number of years
and the procedure when actual data for $n$ years are used with no
scaling. It shows the $\Delta\chi^2$ obtained when all parameters
are kept fixed (for convenience) and the ``theory" is generated with
$\sin^2\theta_{23}^{input}=0.25$, as a function of the number of years
of exposure in ICAL. The smooth red line corresponds to the sensitivity
obtained when 1000 years of events are taken and scaled to the required
number of years ($1,2, \cdots, 20$) to generate the ``data". The green
histogram, in contrast, corresponds to the case when ``data" are generated
for the exact number of years required and then compared to the (scaled)
``theory". It can be seen that the trend of the two lines is the same,
and the $\Delta\chi^2$ value in the second case fluctuates about the
median red line obtained with scaling. This validates our use of the
scaling procedure in our analysis.

\subsection{Effect of errors in hadron energy reconstruction}
\label{ssec:eh}

It can be seen that the primary sensitivity to tau events over the
NC background occurs because the former correspond to higher hadron
energies. Hence the results obtained in the previous sections are
dependent on the correct hadron energy reconstruction. We examine
briefly here the effect of errors in the hadron energy reconstruction.

One of the important issues in the analysis is the hadron energy
reconstruction. ICAL has rather poor sensitivity to the
hadron energy, compared to its ability to reconstruct muons. In our
analysis, the hadron energy of the events generated by the NUANCE
generator is smeared {\em event-by-event} and binned appropriately in
the observed hadron energy bins. In order to understand the dependence
of the sensitivity on the hadron energy, we have therefore simulated
the following: the ``data" is generated according to the ``true" hadron
energy reconstruction determined by the simulations group of the
collaboration \cite{Devi:2013wxa}; the ``theory" however, is generated
using a different width, while retaining the correct central value.

Three sets of widths were used, with $\sigma/E$ of the reconstructed
hadron energy distribution being 5\%, 10\%, and 50\% worse than the
true value\footnote{The original hadron energy response
\cite{Devi:2013wxa} fitted the number of ``hits" in the event to a Vavilov
distribution with 4 fit parameters. Here, for simplicity, we have used
an equivalent Gaussian distribution with only two parameters, mean and
$\sigma$ (excluding overall normalisation in each case) and changed the
latter to simulate the worse results. Since the Vavilov has a longer tail
than the corresponding Gaussian distribution, we have therefore ignored
higher energy tails that would actually improve the result we obtain.}. As
a consequence, some of the events which would have been binned in a
given energy bin may now be binned in any of the adjacent bins.
Since there are more events at low enery, especially below 6 GeV, as
can be seen from  Fig.~\ref{fig:ncandcc}, the application of a worse
reconstruction width for the hadron energy causes more of the low
energy events to smear into even lower energies (and hence are lost to
the analysis if the reconstructed energy is less than 1 GeV), or into
the higher energy, $E_H > 6$ GeV, bins.  This results in the higher
energy bins having a larger number of events than the ``data" set, even
when the oscillation parameters are not changed.  The dominant events
at low energies are NC events, which are independent of oscillations;
the mis-match caused by this error in hadron energy reconstruction can
only be compensated by changing the CC-$\tau$ events, using a different
set of oscillation parameters; this change has to be quite large due to
the smaller number of CC-$\tau$ events compared to the NC sample.

\begin{figure}[thp]
\centering
\includegraphics[width=0.49\textwidth]{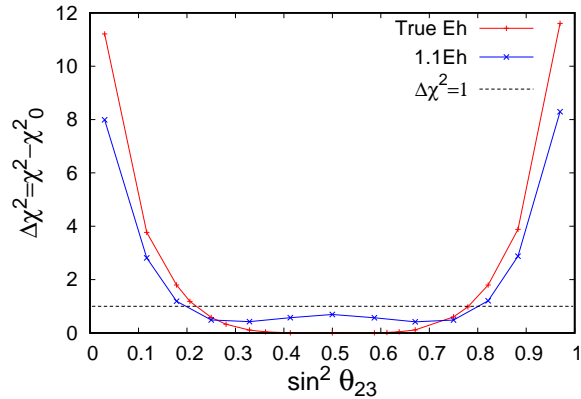}
\caption{The dependence on the hadron energy response. The figure shows
the change in the sensitivity to $\sin^2\theta_{23}$ when the ``theory"
events are smeared by a hadron reconstruction code that has 10\% larger
width than ``data". See text for details.}
\label{fig:wrongeh}
\end{figure}

In Fig.~\ref{fig:wrongeh}, we compare the sensitivity to
$\sin^2\theta_{23}$  when the ``true"
hadron energy response is used for both the ``data" and ``theory" with
that when the latter has a 10\% larger width than the former. Here the
sensitivity is defined as the difference in $\chi^2$:
\begin{equation}
\Delta \chi^2 = \chi^2 - \chi^2_{0}~,
\label{eq:delchi2_Eh}
\end{equation}
where $\chi^2_0$ is the minimum value of $\chi^2$ when both ``data"
and ``theory" use the same hadron energy reconstruction.  While the
sensitivity worsens a little, and the minimum $\Delta \chi^2$ is no longer
at the true value (see below for more details on this), there is still
consistency of the results at $1\sigma$, {\it i.e.}, $\Delta \chi^2 =1$.

\begin{figure}[thp]
\centering
\includegraphics[width=0.49\textwidth]{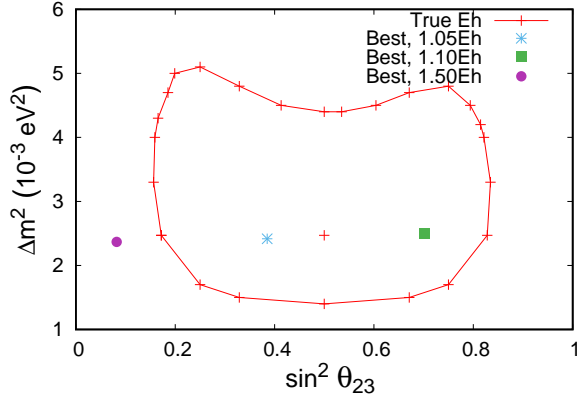}
\caption{70CL contour plot of allowed parameter range in the
$\sin^2\theta_{23}$--$\Delta m^2$ plane with the best fit value marked
with a `plus'. The best fit values when the ``theory" events are
reconstructed with hadron energy resolution having 5\%, 10\% and 50\%
larger widths are also shown as cross, square and circled points. See
text for details.}
\label{fig:contour}
\end{figure}

Fig.~\ref{fig:contour} shows the two dimensional contour plot of allowed
values in $\sin^2\theta_{23}$ and $\Delta m^2$ at the 70\% CL along with
the best fit value (shown as a plus). The best fit values when the
``theory" events are reconstructed according to a hadron energy
response with a width 5\%, 10\% and 50\% larger than that used for the
``data", are also shown (as a cross, a filled square and a filled circle
respectively).

We see that the effect of mismatch in true and fitted hadron energy
response is higher in $\sin^2\theta_{23}$ compared to $\Delta m^2$
as the deviation in the best fit value is rather small for the latter.
Both 5\% and 10\% worse widths give acceptable central values,
with an acceptable minimum $\Delta \chi^2$ ($< 0.4$). When the width used for
``theory" is as much as 50\% worse than for the ``data", the best fit
point lies well outside the contour. In addition, the minimum $\chi^2$
in this case is $\Delta\chi^2_{min} = 12.4$, which is nearly $3.5\sigma$
worse than when the correct hadron energy reconstruction is used for both
``data" and ``theory".

This is an important consideration in the tau neutrino analysis. The
mini-ICAL prototype has been used \cite{ApoorvaDAE} to validate the
energy and momentum response of the detector to muons. However, there is
no data as yet to validate the hadron simulations results that have been
obtained so far, although the simulations themselves have been validated
against data from the MONOLITH experiment \cite{Mohan:2014qua}. A more
detailed analysis using bin-to-bin correlations will allow for a detailed
study of the effect of systematic errors in the hadron energy response
and will improve the quality of the tau neutrino analysis. At this
point, the short summary is that the analysis will tolerate about 10\%
deviations from the expected hadron energy response and such a detector
will continue to be sensitive to these events.

\subsection{Combined study of tau and ``standard" muon events}
So far we have examined the sensitivity of pure hadron events (including
NC and CC-tau events) in ICAL to the neutrino oscillation parameters.
Notice that these arise from the {\em same} atmospheric neutrino fluxes
as the standard muon events in ICAL that are the main goal of this
detector \cite{ICAL:2015stm}.
While the unoscillated muon neutrino fluxes give rise to
the dominant CC-muon events in the detector via $P_{\mu\mu}$, the electron
neutrino fluxes also contribute to this signal via $P_{e\mu}$. Hence,
uncertainties such as overall flux normalisation, zenith angle dependence,
and energy tilt error are the same for both sets of analyses. Since tau
is massive, the CC-tau cross section is highly suppressed at smaller
energies, $E_\nu \lesssim 5$ GeV. However, due to the larger threshold,
tau production events dominantly arise from deep inelastic scattering
and it is reasonable to assume that the cross section uncertainties
(for CC-tau, NC, and CC-mu) are roughly the same. It then becomes
obvious that the systematic uncertainties for the dominant CC-mu processes
are the same as for the CC-tau + NC processes currently being studied.
Since the systematic uncertainties are the dominant factor limiting the
sensitivity in the tau analysis, there is expected to be a significant
improvement on combining the analyses from the two data sets. That is,
given that the two data sets have common systematic uncertainties, the
sensitivity ($\Delta \chi^2$) of the combined analysis is expected to
be better than just the sum of the two individual values.

Previous simulations studies of the
INO collaboration have demonstrated the capability of ICAL with respect
to precision measurement of the 2--3 neutrino oscillation parameters:
$\sin^2\theta_{23}$ and its (as-yet unknown) octant, $\Delta m^2$ (including
its sign), while being insensitive to the CP phase $\delta_{CP}$
\cite{ICAL:2015stm,Mohan:2016gxm}. This
was done using the CC muon events generated when muon neutrinos interact
with the ICAL detector, producing charged muons whose momentum, direction
and sign of charge can be accurately reconstructed from the long tracks
they leave in the (magnetised) detector using a Kalman filter-based
alogrithm \cite{ICAL:2015stm}. Since the tau events indicate (albeit
admittedly limited) sensitivity to these parameters, we now go on to a
combined analysis of these data sets.

The sensitivity to a given neutrino oscillation parameter is again
defined through the $\chi^2$:
\begin{equation}
\chi^2_{\hbox{comb}} = \min_{\xi_{k}}  \left[
\chi^2 (\tau) + \chi^2 (\hbox{std muon}) \right] + 
\chi^2 (\hbox{prior}) + \sum_{k=1}^{N_k} \xi_{k}^{2}~,
\label{eq:chisqnew}
\end{equation}
where the individual contributions are given by
\begin{eqnarray} \nonumber
\chi^{2} (\tau) & = & \sum^{N_{E}}_{i=1} \sum^{N_{cos \theta}}_{j=1} \,
2 \left( \left( T_\tau^{ij} - D_\tau^{ij} \right) -
D_\tau^{ij} \ln \left( \frac{T_\tau^{ij}}{D_\tau^{ij}} \right) \right)~,
\\ \nonumber
\chi^{2} (\hbox{std muon}) & = &  
\chi^{2}_- (\hbox{std muon}) + \chi^{2}_+ (\hbox{std muon})~, \\ \nonumber
\hbox{where } \chi^{2}_\pm (\hbox{std muon}) & = &  \sum^{N_{E_\mu}}_{i=1}
\sum^{N_{cos \theta_\mu}}_{j=1} \sum^{N_{H_\mu}}_{k=1}
\, 2 \left( \left( T_{\mu,\pm}^{ijk} - D_{\mu,\pm}^{ijk} \right) -
D_{\mu,\pm}^{ijk} \ln \left( \frac{T_{\mu,\pm}^{ijk}}{D_{\mu,\pm}^{ijk}}
\right) \right)~,
\end{eqnarray}
and $\chi^2 (\hbox{prior})$ is given by the second term on the right
hand side of Eq.~\ref{eq:prior}, with $\xi_k$ summing over all nuisance
parameters as described earlier. Note that due to the ability to
reconstruct the sign of the muon charge, the contribution from the
individual neutrino and anti-neutrino events are considered for the
standard muon analysis while the summed contribution is analysed for the
tau events. Here the ``tau" contribution is understood to include both
CC-tau and NC events. The terms corresponding to the ``data" and ``theory"
events for the standard muons are analogous to those for the tau events
given in Eq.~\ref{eq:th}, and binned in the three variables, $E_\mu$,
$\cos\theta_\mu$ and $E_{H_\mu}$, corresponding to the muon
momentum magnitude, muon zenith angle, and total hadron energy in the
muon CC event.

\paragraph{Note on coding the systematics}: The $\mu^+$ and $\mu^-$
events are separately analysed and their individual contributions to the
$\chi^2$ determined. The theory contributions to each of these involves
either the set $\xi_-$ or $\xi_+$ of systematic uncertainties; however,
due to a small charge identification inefficiency (less than 2\%
for muon momenta beyond about 1 GeV/c), there is a small fraction of
``wrong-sign" events in each events sample. In the case of the CC-tau+NC
analysis, of course, the neutrino- and anti-neutrino-induced events are
added together. Hence, while solving for the set of $\xi_{k}^{min}$,
a certain simplification is applied: since the anti-neutrino events are
about three times smaller than the neutrino events (due to their
relatively smaller cross sections), terms of the order of $(N^+/N^-)^2$
are dropped from the expressions. While this error is very small for
the case of standard muons, it is about 2--3\% for the case of the
tau events. However, with this approximation, it was found possible to
implement a fast analytical invertor for the corresponding $10 \times 10$
pulls matrix, which speeded up the analysis extensively.

\begin{figure}[thp]
\centering
\includegraphics[width=0.49\textwidth]{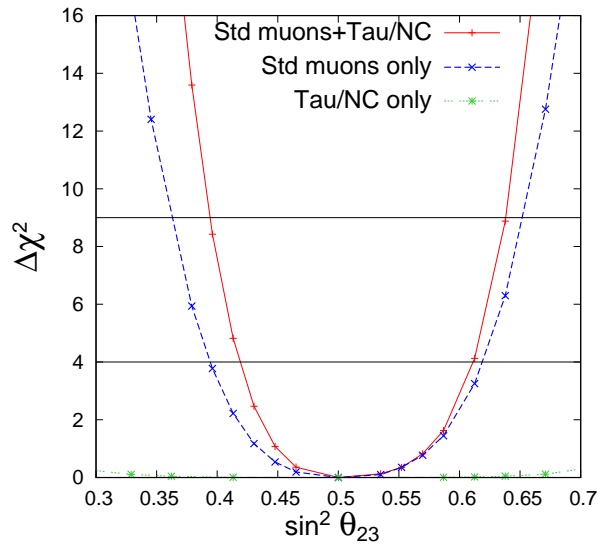}
\caption{10 year sensitivity of ICAL to the oscillation parameter 
$\sin^2\theta_{23}$ with NC + CC tau events alone (this analysis),
from standard muons alone (old analysis \cite{ICAL:2015stm,Mohan:2016gxm}),
and a combined analysis of the two data sets for
$\sin^2\theta_{23}^{in} = 0.5$. It can be seen that while the sensitivity
of tau events to the parameter is very small in the range of values
shown, it causes a dramatic improvement in the precision measurement of
this parameter when combined with the standard muon events.}
\label{fig:combinedt23}
\end{figure}

\paragraph{Effect on sensitivity to $\sin^2\theta_{23}$}: 
Fig.~\ref{fig:combinedt23} shows the effect of combining the tau events
with the standard muon events on the 10 year sensitivity to the
oscillation parameter $\sin^2\theta_{23}$, for the input value
$\sin^2\theta_{23}^{in}=0.5$ ($\theta_{23}^{in} = 45^\circ$).

While the sensitivity of the tau events alone is marginal in the range
of $\sin^2\theta_{23}$ shown in the figure, it significantly improves
the precision reach for this parameter. This is defined as
\begin{equation}
P^{n\sigma} (p) \equiv \frac{\Delta V^p_n}{2V^p_0}~,
\label{eq:prec}
\end{equation}
where $\Delta V_n^p$ is the allowed range of the values of the parameter
$p$ at $n\sigma$, when the remaining parameters are marginalised over
their $3\sigma$ ranges, and $V^p_0$ is its central value. At $2\sigma$,
we see that the precision $P^{2\sigma} ({\sin^2\theta_{23}})$ reduces
from 11\% to 9.5\% on the inclusion of the tau events, and improves even
more significantly from 15\% to 12\% at $3\sigma$. This is because the
{\em source} of atmospheric neutrinos is the same in both cases, and
this helps reduce the effects of including systematic uncertainties,
which are common to both analyses.

\paragraph{Sensitivity to the octant of $\theta_{23}$}: This question is
important for model builders. The dominant contribution to the survival
probabilities such as $P_{\mu\mu}$ is from the octant-insensitive term
$\sin^22\theta_{23}$ and the dependence on the octant-sensitive
$\sin^2\theta_{23}$ is proportional to $\sin^22\theta_{13}$ and
hence is small. There is a larger octant dependence in the oscillation
probabilities ($P_{ij}, i \ne j$), but all such dependences are also
modulated by $\sin^22\theta_{13}$, thus making it challenging to measure.

\begin{figure}
     \centering
     \begin{subfigure}[t]{0.49\textwidth}
         \centering
         \includegraphics[width=\textwidth]{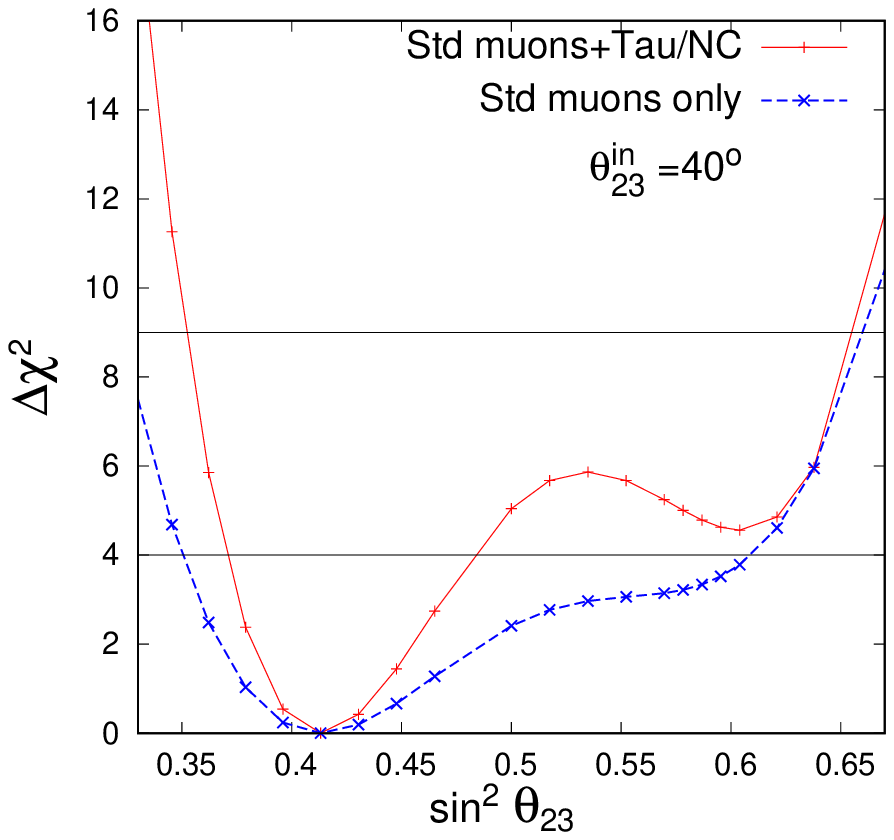}
\caption{ }
         \label{fig:t23_40}
     \end{subfigure}
     \hfill
     \begin{subfigure}[t]{0.49\textwidth}
         \centering
         \includegraphics[width=\textwidth]{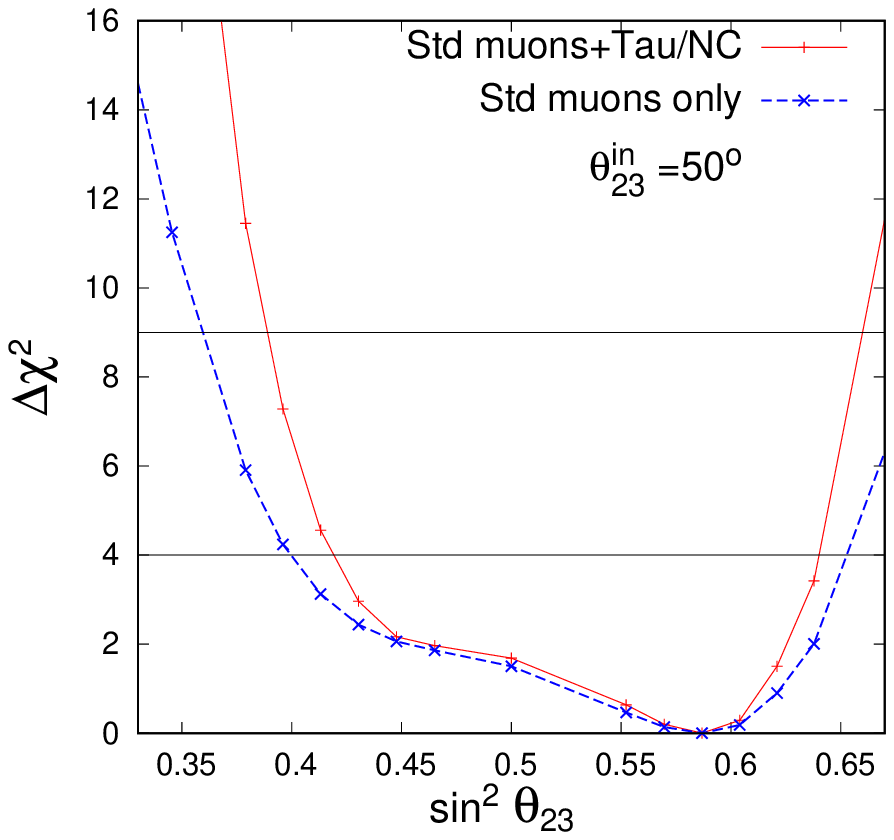}
\caption{ }
         \label{fig:t23_50}
     \end{subfigure}
\caption{10 year sensitivity of ICAL to the octant of the oscillation
parameter $\sin^2\theta_{23}$ from standard muons alone (old analysis
\cite{ICAL:2015stm}), and a combined analysis of the standard muon events
and the tau data sets for input values in the (a) lower and (b) upper
octants with input values $\theta_{23}^{in}=40^\circ$ and $50^\circ$
respectively.}
\label{fig:octant}
\end{figure}

Fig.~\ref{fig:octant} shows the improvement in sensitivity to the octant
of $\theta_{23}$ when tau events are included with the standard muon
analysis. The two plots show the octant sensitivity when input values of
$\theta_{23}^{in}=40^\circ,~50^\circ$ ($\sin^2\theta_{23}^{in}=0.413,~
0.587$) in the lower and upper octants are used. These two sample values
lie well within the $3\sigma$ range of this parameter (for either mass
ordering) as can be seen from Table \ref{tab:osc}.

As seen from Fig.~\ref{fig:t23_40}, there is a significant improvement
in the case of the input value of this parameter being in the lower
octant, {\em viz}., $\theta_{23}^{in}=40^\circ$, with the combined
analysis being able to discriminate against the maximal mixing value
of $\theta_{23}^{in} = 45^\circ$ ($\sin^2\theta_{23}^{in} = 0.5$)
at $2\sigma$. Hence, while the standard muon analysis could not
distinguish the octant (or even deviation from maximality) for the
input value of $\theta_{23}^{in}=40^\circ$, the combined analysis can do
both. As expected, the improvement is more modest for the case when the
input value is in the upper octant, with $\theta_{23}^{in}=50^\circ$
(see Fig.~\ref{fig:t23_50}), due to the nature of the dependence
of $P_{\mu\tau}$ on $\sin^2\theta_{23}$; however, as seen earlier,
the inclusion of the tau events improves the precision to which
$\sin^2\theta_{23}$ can be determined in every case.

\begin{figure}[thp]
\centering
\includegraphics[width=0.49\textwidth]{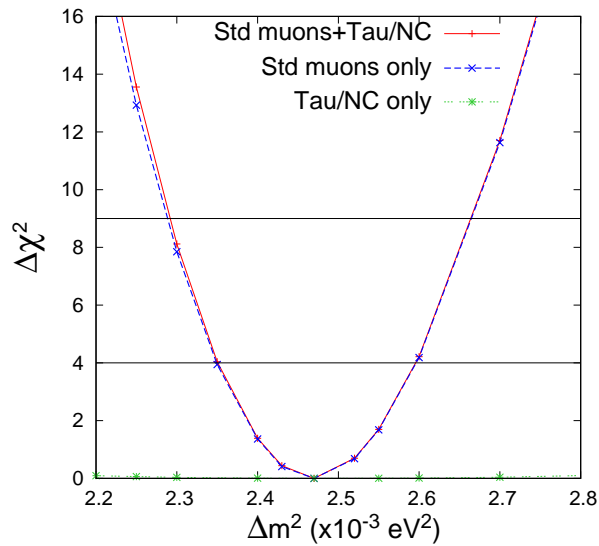}
\caption{Sensitivity to oscillation parameter $\Delta m^{2}$ from
standard muons alone, tau events alone, and from combining the two sets
in a simultaneous analysis.}
\label{fig:combineddelm2}
\end{figure}

\paragraph{Effect on sensitivity to $\Delta m^2$}:
Fig.~\ref{fig:combineddelm2} shows the effect of combining the tau
events with the standard muon events on the 10 year sensitivity to the
oscillation parameter $\Delta m^2$, for the input value $\Delta
m^2_{in} = 2.47 \times 10^{-3}$ eV$^2$. It is seen that the sensitivity on
including the tau events is marginal. Note that the results from
standard muons alone are slightly different from those shown in
Ref.~\cite{Mohan:2016gxm} due to the slightly different central values of
parameters used in the analysis.

\paragraph{Sensitivity to the mass ordering or the sign of $\Delta
m^2$}: The combined standard muon and tau+NC events have a marginally
better sensitivity to the sign of $\Delta m^2$. For instance, the
value of $\Delta \chi^2$ from the combined analysis is 0.5 better
than that obtained with standard muons alone for the central
values listed in Table \ref{tab:osc}, assuming the normal ordering and
10 years' running of ICAL \cite{Mohan:2016gxm}. There is hardly any
improvement if the true hierarchy is assumed to be inverted.

\section{Discussion and Conclusions}

Tau neutrinos do not naturally occur in the atmospheric neutrino
spectrum, which comprises electron and muon neutrinos (and
anti-neutrinos) primarily arising from pion and subsequent muon decays
of the primary cosmic ray spectrum. Due to neutrino oscillations, those
atmospheric neutrinos that traverse significant path lengths can
oscillate into one another as well as into tau neutrinos. Hence it is
expected that a significant fraction of upward-going atmospheric
neutrinos are of the tau flavour. Such tau neutrinos can provide an
independent test of neutrino oscillations through their direct detection
via charged current (CC) interactions of these neutrinos with the
material of the detector. In this paper we have made a detailed
simulations study of such a process at the proposed magnetised ICAL
detector at the India-based Neutrino Observatory, INO. While ICAL
is being optimised to detect charged muons from the CC interactions of
atmospheric muon neutrinos (and anti-neutrinos), the so-called standard
muon sample, it is also sensitive to hadrons that are produced along
with muons in the CC interaction.

In particular, we have analysed the events where the charged taus
produced in the CC interaction decay hadronically so that the event is
rich in hadrons (produced at the interaction vertex as well as during
tau decay). Due to the characteristics of the ICAL detector, such events
are indistinguishable from neutral current (NC) events where the final
state neutrino escapes and only a hadron residue is observable. Hence both
CC-tau events and NC events are analysed together in this work. Since the
tau production threshold is high, $E_\nu > 3.5$ GeV, due to the large mass
of the tau lepton, the tau events are few in number but make a significant
addition over the high energy NC events which are also limited due to
the steeply falling neutrino spectrum ($\propto E_\nu^{-2.7}$).

Another signature of tau events is the fact that they are exclusively
produced by tau neutrinos moving in the upward direction. Since the
neutrino energies involved are sufficiently large, the final state tau
as well as its decay products are also peaked in the same direction
as the initial neutrino. Hence good angular discrimination should
enable extraction of these events over the uniformly distributed NC
events. Due to the limitations of detector reconstruction, the direction
resolution of hadrons is rather poor in ICAL. However, the addition of
just two angle bins (corresponding to UP going and DOWN going events)
leads to a significant sensitivity of these events to the presence of
tau neutrinos. We have found that ignoring the tau component in the
theory fits to the simulated data (which contains both NC and CC-tau
events) leads to a mismatch with $\Delta\chi^2=38$ which drops to 15 when
various systematic errors are included in the analysis, thus indicating
that the pure hadron events sample can unambiguously signal the presence
of tau neutrinos in the atmospheric neutrino flux.

For the first time, we have also analysed the CC-tau + NC sample for its
sensitivity to the neutrino oscillation parameters themselves. A
modest sensitivity to both the 2--3 parameters $\sin^2\theta_{23}$
and $\vert \Delta m^2 \vert$ (although the sample had no sensitivity to
the sign of $\Delta m^2$) was found, mainly limited by the systematic
uncertainties. Although small, the result was encouraging since this data
sample originates from the {\em same} atmospheric neutrino fluxes as the
standard muon sample that has been extensively studied by the INO collaboration \cite{ICAL:2015stm}. Hence systematic unertainties pertaining to the
fluxes, such as overall normalisation, zenith angle, and energy dependent
tilt uncertainties are the same for both samples. In addition, at higher
energies dominated by deep inelastic scattering processes, the cross
section uncertainties can also be considered to be the same for the two
sets. Hence it is possible to perform a combined analysis of the CC-tau+NC
and the standard muon sample thereby increasing the statistics without
worsening the systematic uncertainties. Such a combined analysis gave
a sensitivity ($\Delta \chi^2$) that was significantly better than just
the sum of the individual contributions. In particular, while there was
not much improvement in the precision of $\vert \Delta m^2 \vert$, there
was a significant improvement in the precision of $\sin^2\theta_{23}$. In
addition, it was found that the sensitivity to the octant of $\theta_{23}$
significantly improved. For instance, it is possible to determine both
the octant as well as establish deviation from maximality at $2\sigma$
from a 10 year combined sample when the input value of $\theta_{23}$
was in the first octant, $\theta_{23}^{in} = 40^\circ$. A somewhat more
modest improvement was seen when the input value of $\theta_{23}$ was
assumed to be in the second octant, as expected. A small improvement in
the determination of the neutrino mass ordering was found when the true
ordering was assumed to be normal; no such improvement was seen when the
true ordering was assumed to be inverted.

Neutrino experiments are low counting experiments. Hence it is important
to analyse every possible channel to yield more information on the
neutrino oscillation parameters. Combining tau events with standard muon
events opens up a way of improving the precision and possible
measurement of parameters such as the as-yet unknown octant of the
2--3 oscillation parameter, $\theta_{23}$. Many current and up-coming
experiments are focussing on this relatively unknown sector. In addition,
information from the tau sector can also probe the 3-flavour structure
of the PMNS matrix and unitarity violation \cite{Denton:2021mso}. This
should give even more impetus to the study of such tau events.

\paragraph{Acknowledgements}: We thank M.V.N. Murthy, Jim Libby,
S.M. Lakshmi and the INO collaboration for detailed discussions and
LSM for code validation. We thank the referee for several insightful
suggestions which tremendously improved the content of the manuscript.


\begin{thebibliography}{99}

\bibitem{ICAL:2015stm}
S.~Ahmed \textit{et al.} [ICAL],
Pramana \textbf{88}, no.5, 79 (2017)
doi:10.1007/s12043-017-1373-4
[arXiv:1505.07380 [physics.ins-det]].

\bibitem{Chacko:2019wwm}
A.~Chacko, D.~Indumathi, J.~F.~Libby and P.~K.~Behera,
Phys. Rev. D \textbf{102}, no.3, 032005 (2020)
doi:10.1103/PhysRevD.102.032005
[arXiv:1912.07898 [physics.ins-det]].

\bibitem{Super-Kamiokande:2012xtd}
    Abe, K. et al., 
    [Super-Kamiokande],
    Phys. Rev. Lett. {\bf 110} (2013) 181802, 
    arXiv: 1206.0328, 
    doi:10.1103/PhysRevLett.110.181802.



\bibitem{IceCube:2019dqi}
M.~G.~Aartsen \textit{et al.} [IceCube],
Phys. Rev. D \textbf{99}, no.3, 032007 (2019)
doi:10.1103/PhysRevD.99.032007
[arXiv:1901.05366 [hep-ex]].

\bibitem{Denton:2021rsa}
P.~B.~Denton,
Phys. Rev. D \textbf{104}, no.11, 113003 (2021)
doi:10.1103/PhysRevD.104.113003
[arXiv:2109.14576 [hep-ph]].

\bibitem{DeGouvea:2019kea} 
A.~De Gouv\^ea, K.~J.~Kelly, G.~V.~Stenico and P.~Pasquini,
Phys. Rev. D \textbf{100}, no.1, 016004 (2019)
doi:10.1103/PhysRevD.100.016004
[arXiv:1904.07265 [hep-ph]].

\bibitem{KM3NeT:2021ozk}
S.~Aiello \textit{et al.} [KM3NeT],
Eur. Phys. J. C \textbf{82}, no.1, 26 (2022)
doi:10.1140/epjc/s10052-021-09893-0
[arXiv:2103.09885 [hep-ex]].

\bibitem{Athar:2005wg}
    Athar, H. Kim, C. S. and Lee, Jake,
    Mod. Phys. Lett. \textbf{A 21} (2006) 1049,
    arXiv: hep-ph/0505017,
    doi: 10.1142/S021773230602038X.

\bibitem{Pasquali:1998xf}
L.~Pasquali and M.~H.~Reno,
Phys. Rev. D \textbf{59}, 093003 (1999)
doi:10.1103/PhysRevD.59.093003
[arXiv:hep-ph/9811268 [hep-ph]].

\bibitem{Athar:2001jw}
H.~Athar, K.~M.~Cheung, G.~L.~Lin and J.~J.~Tseng,
Astropart. Phys. \textbf{18}, 581-592 (2003)
doi:10.1016/S0927-6505(02)00183-4
[arXiv:hep-ph/0112222 [hep-ph]].

\bibitem{Lee:2004zm}
F.~F.~Lee and G.~L.~Lin,
Astropart. Phys. \textbf{25}, 64-73 (2006)
doi:10.1016/j.astropartphys.2005.11.003
[arXiv:hep-ph/0412383 [hep-ph]].

\bibitem{Casper:2002sd}
D.~Casper,
Nucl. Phys. B Proc. Suppl. \textbf{112}, 161-170 (2002), 
doi:10.1016/S0920-5632(02)01756-5, 
[arXiv:hep-ph/0208030 [hep-ph]].

\bibitem{GEANT4:2002zbu}
    Agostinelli, S. et al,
    [GEANT4],
    Nucl. Instrum. Meth. \textbf{A 506} (2003) 250,
    doi: 10.1016/S0168-9002(03)01368-8.
\bibitem{Honda:2006qj}
M.~Honda, T.~Kajita, K.~Kasahara, S.~Midorikawa and T.~Sanuki,
Phys. Rev. D \textbf{75}, 043006 (2007)
doi:10.1103/PhysRevD.75.043006
[arXiv:astro-ph/0611418 [astro-ph]].


\bibitem{Honda:2011nf}
M.~Honda, T.~Kajita, K.~Kasahara and S.~Midorikawa,
Phys. Rev. D \textbf{83}, 123001 (2011)
doi:10.1103/PhysRevD.83.123001
[arXiv:1102.2688 [astro-ph.HE]].

\bibitem{Honda:2019ymh}
M.~Honda, M.~Sajjad Athar, T.~Kajita, K.~Kasahara and S.~Midorikawa,
Phys. Rev. D \textbf{100}, no.12, 123022 (2019)
doi:10.1103/PhysRevD.100.123022
[arXiv:1908.08765 [astro-ph.HE]].

\bibitem{Wolfenstein:1977ue}
    Wolfenstein, L., 
    Phys. Rev. \textbf{D 17} (1978) 2369,
    doi: 10.1103/PhysRevD.17.2369.

\bibitem{ParticleDataGroup:2020ssz}
P.~A.~Zyla \textit{et al.} [Particle Data Group],
PTEP \textbf{2020}, no.8, 083C01 (2020)
doi:10.1093/ptep/ptaa104

\bibitem{Indumathi:2006gr}
D.~Indumathi, M.~V.~N.~Murthy, G.~Rajasekaran and N.~Sinha,
Phys. Rev. D \textbf{74}, 053004 (2006)
doi:10.1103/PhysRevD.74.053004
[arXiv:hep-ph/0603264 [hep-ph]].

\bibitem{Esteban:2020cvm}
I.~Esteban, M.~C.~Gonzalez-Garcia, M.~Maltoni, T.~Schwetz and A.~Zhou,
JHEP \textbf{09}, 178 (2020)
doi:10.1007/JHEP09(2020)178
[arXiv:2007.14792 [hep-ph]].

\bibitem{Dziewonski:1981xy}
A.~M.~Dziewonski and D.~L.~Anderson,
Phys. Earth Planet. Interiors \textbf{25}, 297-356 (1981)
doi:10.1016/0031-9201(81)90046-7

\bibitem{Mohan:2016gxm}
L.~S.~Mohan and D.~Indumathi,
Eur. Phys. J. C \textbf{77}, no.1, 54 (2017)
doi:10.1140/epjc/s10052-017-4608-0
[arXiv:1605.04185 [hep-ph]].

\bibitem{Devi:2013wxa}
M.~M.~Devi, A.~Ghosh, D.~Kaur, L.~S.~Mohan, S.~Choubey, A.~Dighe, D.~Indumathi, S.~Kumar, M.~V.~N.~Murthy and M.~Naimuddin,
JINST \textbf{8}, P11003 (2013)
doi:10.1088/1748-0221/8/11/P11003
[arXiv:1304.5115 [physics.ins-det]].

\bibitem{Datta:2021myx}
J.~Datta, B.~Singh and S.~U.~Sankar,
[arXiv:2111.14184 [hep-ph]].

\bibitem{Indumathi:2009hg}
D.~Indumathi and N.~Sinha,
Phys. Rev. D \textbf{80}, 113012 (2009)
doi:10.1103/PhysRevD.80.113012
[arXiv:0910.2020 [hep-ph]].

\bibitem{Devi:2018ltf}
M.~M.~Devi, A.~Dighe, D.~Indumathi and S.~M.~Lakshmi,
JINST \textbf{13}, no.03, C03006 (2018)
doi:10.1088/1748-0221/13/03/C03006

\bibitem{Blennow:2013oma}
M.~Blennow, P.~Coloma, P.~Huber and T.~Schwetz,
JHEP \textbf{03}, 028 (2014)
doi:10.1007/JHEP03(2014)028
[arXiv:1311.1822 [hep-ph]].

\bibitem{Cowan:2010js}
G.~Cowan, K.~Cranmer, E.~Gross and O.~Vitells,
Eur. Phys. J. C \textbf{71}, 1554 (2011)
[erratum: Eur. Phys. J. C \textbf{73}, 2501 (2013)]
doi:10.1140/epjc/s10052-011-1554-0
[arXiv:1007.1727 [physics.data-an]].

\bibitem{Kameda}
  J.~Kameda, Ph.D.\ thesis,
  \verb"http://www-sk.icrr.u-tokyo.ac.jp/doc/sk/pub/".
   
%
\bibitem{Ishitsuka} 
  M.~Ishitsuka,  Ph.D.\ thesis,
  \verb"http://www-sk.icrr.u-tokyo.ac.jp/doc/sk/pub/".                   

\bibitem{Super-Kamiokande:2017edb}
Z.~Li \textit{et al.} Super-Kamiokande collab.,
Phys. Rev. D {\bf 98}, no.5, 052006 (2018), 
doi:10.1103/PhysRevD.98.052006
[arXiv:1711.09436 [hep-ex]].

\bibitem{ApoorvaDAE}
Bhatt, A.D., Majumder, G., Datar, V.M., Satyanarayana, B. (2021). Muon
Momentum Spectra with mini-ICAL. In: Behera, P.K., Bhatnagar,
V., Shukla, P., Sinha, R. (eds) XXIII DAE High Energy Physics
Symposium. Springer Proceedings in Physics, vol 261. Springer,
Singapore. https://doi.org/10.1007/978-981-33-4408-2\_102.

\bibitem{Mohan:2014qua}
L.S. Mohan et al., JINST {\textbf 9} (2014) 09, T09003
        [arXiv:1401.2779 [physics.ins-det]].

\bibitem{Denton:2021mso}
P.~B.~Denton and J.~Gehrlein,
[arXiv:2109.14575 [hep-ph]], 2021.



\end{thebibliography}
\end{document}